\documentclass{aa} 

\usepackage{graphicx}
\usepackage{tabularx}
\usepackage{txfonts}
\newcommand{\ks}{{$\rm K_s$}\ }

\newcommand{\jks}{{$\rm J-K_s$}\ }
\newcommand{\ejks}{{$\rm E(J-K_s)$} }

\begin{document}


   \title{Observed kinematics of the Milky Way nuclear stellar disk region}


   \author{M. Zoccali\inst{1,2}
          \and
          A. Rojas-Arriagada\inst{3,2,5,6}
          \and
          E. Valenti\inst{4,7}
          \and
           R. Contreras Ramos\inst{2}
           \and
          A. Valenzuela-Navarro\inst{1,2}
           \and
          C. Salvo-Guajardo\inst{1}
          \fnmsep\thanks{Based on observations taken within the ESO VISTA Public Survey VVV, Program ID 179.B-2002.}
          }

   \institute{Instituto de Astrof\'isica, Pontificia Universidad Cat\'olica de Chile, Av. Vicu\~na Mackenna 4860, 782-0436 Macul, Santiago, Chile\\
              \email{mzoccali@astro.puc.cl}
         \and
            Millennium Institute of Astrophysics, Av. Vicu\~na Mackenna 4860, 82-0436 Macul, Santiago, Chile 
         \and
            Departamento de F\'isica, Universidad de Santiago de Chile, Av. Victor Jara 3659, Santiago, Chile
         \and
            European Southern Observatory, Karl Schwarzschild-Strabe 2, 85748 Garching bei München, Germany
        \and
        N\'ucleo Milenio ERIS
        \and
        Center for Interdisciplinary Research in Astrophysics and Space Exploration (CIRAS), Universidad de Santiago de Chile, Santiago, Chile
        \and
              Excellence Cluster ORIGINS, Boltzmann\--Stra\ss e 2, D\--85748 Garching bei M\"{u}nchen, Germany
          }

   \date{Received September 15, 1996; accepted March 16, 1997}

 
  \abstract
   {The nuclear region of the Milky Way, within approximately $-$1$^\circ$$<$$l$$<$+1$^\circ$ and $-$0.3$^\circ$$<$$b$$<$+0.3$^\circ$ (i.e., $|l|$$<$150 pc, $|b|$$<$45 pc), is believed to host a nuclear stellar disk, co-spatial with the gaseous central molecular zone. Previous kinematical studies detected 
   faster rotation for the stars belonging to the nuclear stellar disk, compared to the surrounding regions.}
   {We analyze the rotation velocity of stars at the nuclear stellar disk, and compare them with its analog in a few control
   fields just outside this region. We limit our analysis to stars in the red clump of the color magnitude diagram, in order to be able to relate their mean de-reddened luminosity with distance along the line of sight.}
   {We used a proper motion catalog, obtained from point spread function photometry on VISTA variables in the V\'\i a L\'actea images, to construct maps of the transverse velocity for these stars. We complemented our analysis with radial velocities from the 17$^{th}$ data release of the APOGEE survey.}
   {We find that the main difference between the nuclear stellar disk region and its surroundings is that at the former we see only stars moving eastward, which we believe are located in front of the Galactic center. On the contrary, in every other direction, we see
   the brightest red clump stars moving eastward, and the faintest ones moving westward, as expected for a rotating disk. We interpret these observations as being produced by the central molecular zone, hiding stars behind 
   itself. What we observe is compatible with being produced by just the absence of the component at the back, without requiring the presence of a cold, fast rotating disk. This component is also not clearly detected in the newest release of the APOGEE catalog. In other words, we find no clear signature of the nuclear stellar disk as a distinct kinematical component.
   }
   {This work highlights the need for nearby control fields when attempting to characterize the 
   properties of the nuclear stellar disk, as the different systematics affecting this region, compared
   to nearby ones, might introduce spurious results. Deep, wide field and high resolution photometry of the inner 4 degrees of the Milky Way is needed in order to understand the structure and kinematics of this very unique region of our Galaxy.}
   

   \maketitle
%

\section{Introduction}

The innermost $\sim$2 square degrees ($|l|$$<$1$^\circ$, $|b|$$<$0.3$^\circ$) are the most interesting and yet the least understood region of the Milky Way (MW). They host the nuclear stellar disk (NSD), the central molecular zone (CMZ), three massive young clusters, several powerful X-ray sources, radio bubbles, and the supermassive black hole \citep[see][for a recent review]{bryant+krabbe2021}. Each of these components is rather unique in our Galaxy, and the links among them, in terms of origin and evolution, are very poorly understood. Yet, this is not an exotic and rare region: every spiral galaxy has a central region with similar components. Also, there are strong reasons to believe that what happened in the MW center, in the past, has influenced the evolution of, at least, the whole bulge \citep[e.g.,][for a recent review]{croton2006, laha+2021}.  Therefore, this region definitely deserves to be extensively explored, and the interplay among its various components understood.

An early review about the nuclear region of our Galaxy was offered by \citet{mezger+1996}. They defined a "Nuclear bulge" as being contained within R$<$300 pc from the MW dynamical center. \citet{launhardt+2002} emphasized how this component is distinct from the main bulge, at 300$<$R$<$3000 pc because it shows a sharp increase in the stellar density, it hosts a relatively large population of massive stars \citep[e.g.,]{dong+2012} and star clusters \citep[][and references therein]{hosek+22, schoedel+20, schoedel+23}, and it contains a large fraction of the MW dense molecular gas: the CMZ. \citet{schonrich+2015} presented a first kinematical detection of the NSD as a flat, fast rotating (V$_{\rm rot}$=120 km/s) component with a radius R$_{\rm NSD}$=150 pc. A smaller radius (R$_{\rm NSD}$$\sim$90 pc) has been inferred by \citet{gallego-cano+20} from an analysis of {\it Spitzer}/InfraRed Array Camera (IRAC)  4.5  $\mu$m images. Also, a change in the orientation of the main bar in the inner region has been reported by \citet{gonzalez+11}, who suggested the presence of an inner bar (or a ring) with a semi-major axis of $\sim$500 pc. In addition to \citet{schonrich+2015}, spectroscopic studies of this region have been scarce due to the extreme extinction and stellar crowding \citep[e.g.,][]{surot19phot, surot20ext}. One exception to this is the KMOS spectroscopic survey of \citet{Fritz2021}, which was designed to cover the whole range in age and metallicity over the whole purported NSD area. Based on this spectroscopic sample of about 3100 K/M giant stars, \citet{schultheis+21} studied the kinematics and global metallicity of the NSD, finding evidence for a dynamically cool and metal-rich component of it, whose rotation curve is similar to that derived for the gas component from molecular gas tracers.

More recently, the NSD has been the main target of the GALACTICNUCLEUS program \citep{nogueras-lara+19}. Based on the luminosity function of red clump (RC) stars, the authors found that: {\it i)} the NSD has formed $\sim$95$\%$ of its mass about 8 Gyr ago, while the other $\sim$5$\%$ was formed in a starburst event $\sim$1 Gyr ago \citep{nogueras-lara+20}, with a possible radial age gradient \citep{nogueras-lara+23}. They provided extinction maps and dereddened photometry in \citet{nogueras-lara+2021}. The same group derived proper motions (PMs) for RC stars in the NSD from different datasets. In \citet{martinez-arranz+22}, they compared the GALACTICNUCLEUS astrometry with that of \citet{zoccali+2021}, and identified a low reddened population that shows positive mean longitudinal PM (i.e., moves eastward) and a high reddening one that shows negative mean longitudinal PM. In \citet{shahzamanian+22}, instead, they used as a first  epoch the HST Paschen alpha photometry collected by \citet{wang+2010}, to confirm the detection of two populations moving eastward and westward, in longitude, in addition to a larger population with $\mu_l$ centered at zero that they interpreted as being due to the main bulge. Finally, in \citet{nogueraslara+22los}, they used the PMs by \citet{libralato+21} to confirm the detection of the two components with an opposite $\mu_l$, and also point out that the one moving eastward is 0.1 mag brighter than the other. This is consistent with a radius of the NSD of $R_{\rm NSD}$=150 pc.

The radial velocity data by \citet{Fritz2021}, coupled with PM from the VIRAC2 catalog \citep[an updated version of][]{smith+18VIRAC}, were used by \citet{sormani+22} to constrain a dynamical model of the NSD. The model has a total mass of M$_{\rm NSD}$=10.5$\times$10$^8$ M$_\odot$, a scale length of R$_{\rm NSD}$=88.6 pc, and a velocity dispersion of $\sim$70 km/s.

The star formation history of the NSD is a particularly interesting topic as it defines the interplay between this stellar component and the CMZ, which are both located roughly in the same region \citep{molinari+2011}. The dense molecular gas in the CMZ fulfills the conditions to be the site of a significant star formation, according to our understanding of the surface density of dense molecular gas being the main driver of star formation. Instead, very little star formation is present only in a few of the densest molecular clouds, amounting to one order of magnitude lower than expected for the total CMZ \citep{henshaw+2022, sormani+20, barnes+2017}. Whether this has been different in the past, as suggested by some simulations \citep{armillotta+19}, is something that only the age distribution of NSD stars can tell us, but the current data allow only rather qualitative analyses \citep{matsunaga+11, nogueras-lara+20}.

Specifically, the star formation history of the NSD was derived by \citet{nogueras-lara+20} from the analysis of a dereddened luminosity function reaching $\sim$2 magnitude below the RC. The authors explain that the vast majority of the stars in this LF belong to the RGB+RC. As is well known, the shape of the RC and the RGB is only very weakly affected by age. As has been shown by 
\citet[][their Fig.5c]{girardi+16} that the absolute \ks mag of RC stars, for a solar metallicity population, is indeed constant for ages larger than 2 Gyr, and it is $\sim$0.4 mag fainter than that, only for stars in the small age range close to 1 Gyr, where it basically merges with the RGB bump. Moreover, the same RC feature was also used to derive the extinction law, the reddening, and the distribution of the stars along the line of sight. This can cause important degeneracies, hence deriving independent information about the age distribution, by using variable stars \citep[e.g.,][]{sanders+22} or about the 3D structure and kinematics, by means of PMs and RVs is especially important to reduce the number of unknown parameters.

In this paper we use point spread function (PSF) photometry and PMs from the VISTA Variables in the V\'\i a L\'actea survey \citep[hereafter VVV][]{minniti+2010} to explore the kinematics of the MW nuclear region, with a special emphasis on the NSD.
We also repeated the RV analysis by \citet{schonrich+2015} using the latest data release of APOGEE, finding important differences with respect to their paper.
\section{The catalog} \label{data}

   \begin{figure}
   \centering
   \includegraphics[width=\hsize]{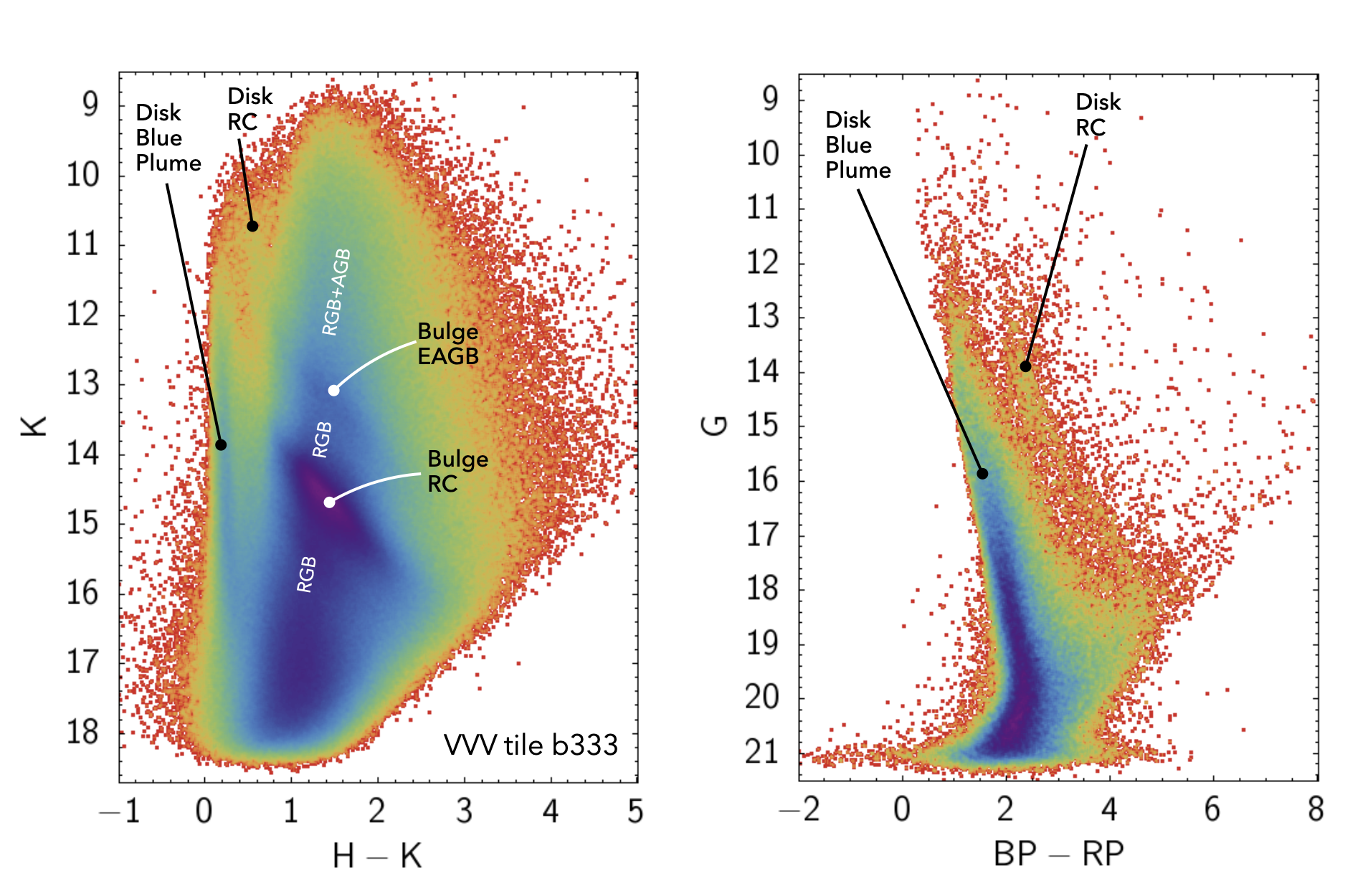}
   \caption{CMD of the central region of the present catalog (tile b333) as seen in VVV (left) and in Gaia DR3 (right). The K magnitude is always the \ks band of VVV calibrated to 2MASS (see text), here labeled simply K to avoid too many subindices. The main features of both diagrams are labeled here and explained in the text.}
   \label{Fig:CMD_VVV_Gaia}
    \end{figure}

For the present analysis we use multi-epoch PSF photometry and PMs derived with the method described in \citet{contreras+17}.
Figure\ref{Fig:CMD_VVV_Gaia} shows the CMD of the region mapped by tile b333 (i.e., $-$1.25$<$$l$$<$0.2 and
$-$0.45$<$$b$$<$0.65) as seen by VVV (left) and by Gaia DR3. The K magnitude is always the \ks band of VVV calibrated to 2MASS (see below), hereafter labeled simply K or K$_0$ in order to avoid too many subindices in the figures.
The CMD from VVV is dominated by bulge evolved stars, including the almost vertical Red Giant Branch  (RGB) and a 
prominent RC which extends diagonally, due to the large dispersion in extinction of this region. Also visible are the Early Asymptotic Giant Branch (EAGB), parallel to the RC and brighter, but much less populated.
A trained eye can also see the RGB bump, not marked in the figure to avoid saturation, as a second diagonal clump about 0.4 mag fainter than the RC. The disk foreground, instead, makes up the two almost vertical sequences on
the left side of this CMD, that is, the Main Sequence (Blue Plume) and a RC, both of them spread at different distances and reddenings. As a small optical telescope, Gaia sees only the bluest features of this CMD, i.e., the disk foreground.
There is only a hint of a RGB at the bottom-right, but it is not well defined, and PMs for these stars are
obviously not precise. No selection against blended stars is made on this CMD from Gaia, therefore it also shows
poorly measured stars at the bottom.
The b333 tile shown here is roughly centered on the GC (see Fig.~\ref{Fig:density}) therefore it is the most crowded and reddened. As we move out from this region the completeness increases, the limit magnitude moves fainter, and the color spread decreases due to lower spread in reddening.

We focus here only on RC stars, for which the photometry is virtually identical to the one described and released in \citet{surot19phot}, except for the fact that the latter provides magnitudes only for the J and \ks bands, while the present ones includes the H band as well. Across the whole b318 tile, representative of the analyzed region, the mean magnitude differences for all the stars in common, without selections, are 0.073 and  0.077 mag, while the inter-quartile ranges are 0.16 and 0.13 mag, for the J and \ks band, respectively. These small differences can be attributed to the fact that our catalogs are corrected from a known zero-point problem of the CASU VVV photometry (see below).
Both catalogs have been obtained by means of the DAOPHOT/ALLSTAR codes of stellar photometry \citep{daophot}, but \citet{surot19phot} also used ALLFRAME \citep{allframe} in order to reach as deep as possible, while \citet{contreras+17} omitted this step in order to keep the astrometry on each frame as independent as possible from the others, to optimize PM measurements. As a result, the limit magnitude and completeness of the \citet{surot19phot} catalog are slightly different, and the present catalog includes PMs. We interpolated \ejks extinction values from the map of \citet{surot20ext} for each star in our catalog. We note that the differences in \jks between the present catalog and \citet{surot20ext} is 0.004 mag, therefore those reddenings are perfectly applicable to our magnitudes, in spite of the small differences mentioned above.

\subsection{Photometric calibration}

The photometry has been calibrated to the VVV photometric system following an improvement over the prescriptions by \citet{hajdu+20}. Briefly: isolated stars in common with the 2MASS catalog \citep{skrutskie+06} were identified within each detector of the VVV mosaic. In order to avoid the many blends in the 2MASS catalog, and taking into account the typical FWHM of stars in those images, only stars separated by at least 3 arcsecs from their neighbors were used for this match.
Their magnitudes were transformed to the VVV photometric system by means of the color and extinction terms reported in \citet{gf+18_VVVcal}, and then a zero point was derived, and applied, as the median difference between the VVV instrumental magnitudes and the transformed 2MASS magnitudes, for each band. It is worth mentioning that this calibration was performed with special attention to both the extreme crowding, blending many stars in 2MASS, and saturation, affecting the brightest stars in VVV. The first problem was minimized by selecting only common stars that did not have any companion, in VVV, within 3 arcsec: the reported typical Full Width at Half Maximum of 2MASS. As for the saturation, we verified that the presence of saturation could be automatically detected in a plot of the magnitude difference (VVV-2MASS) as a function of the 2MASS magnitude, for each band. Therefore, we implemented bright cuts, to ensure that the common stars were not affected by saturation. The number of common stars finally used for each detector ranges between $\sim$20 and 300, in the region analyzed here. The associated error in the calibration was estimated as the MAD of the distribution, as recommended by \citet[][their Sec.~4.5]{gonzalez-fernandez+18}, and \citet{hajdu+20}, in order to take into account both the statistical and systematic contributions, such as the conversion from the 2MASS to the VIRCAM filter set. Typical errors in the region analyzed here are in the range 0.02-0.1 mag in the \ks band, and 0.04-0.07 mag in the H band. Although these errors are non negligible, they are very small compared to the magnitude range used to select RC stars. In addition, all the conclusions drawn here are based on global trends for stars $\sim$0.5 mag brighter/fainter than the RC, therefore, even an error of 0.1 mag would not affect our results.

   \begin{figure}
   \centering
   \includegraphics[width=\hsize]{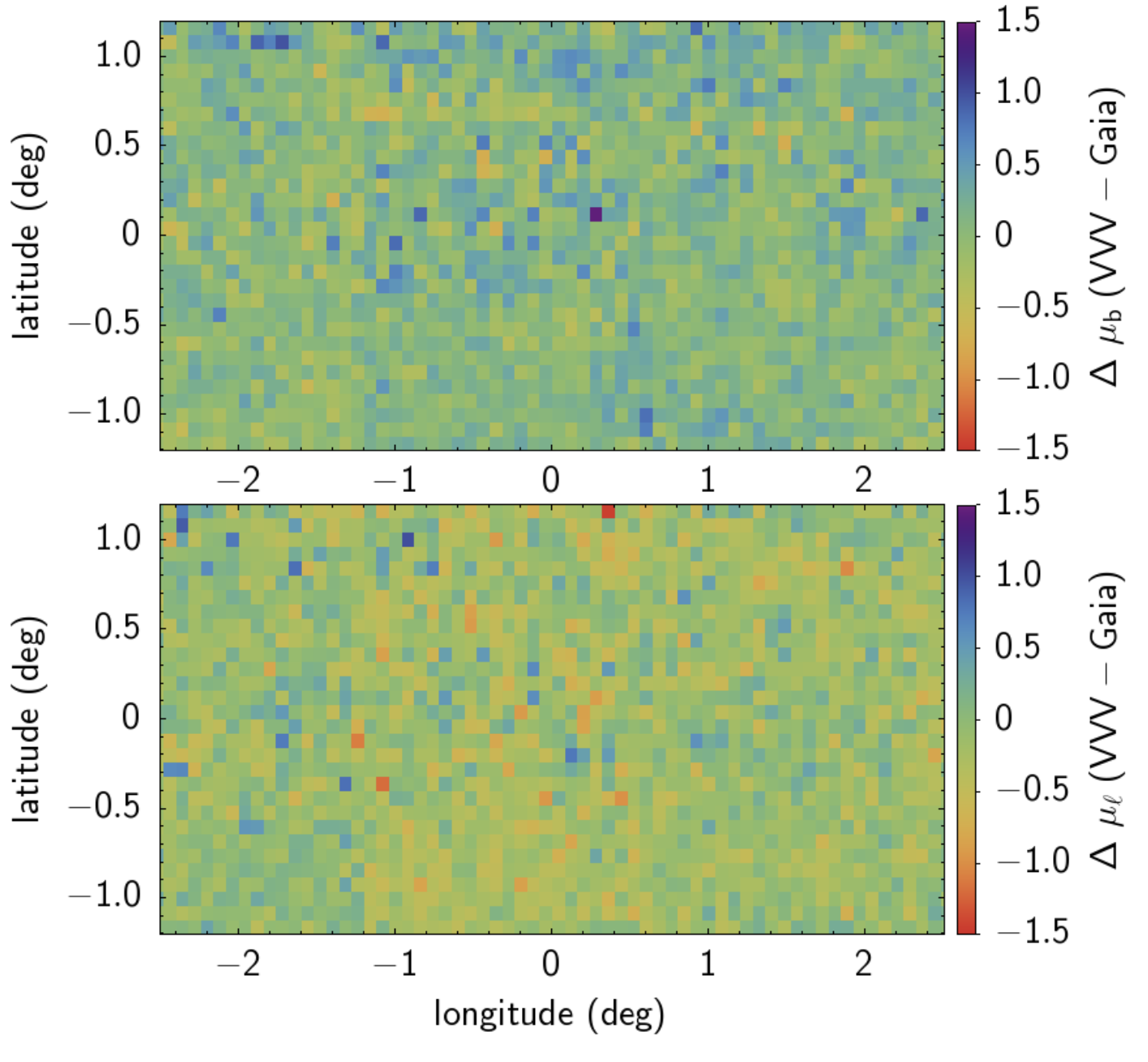}
   \caption{Difference between the PM of individual stars in our catalog versus Gaia DR3, in latitude ({\it Top}) and longitude ({\it bottom}). These maps are based on the stars in common between VVV and Gaia, which are also the ones used for the astrometric calibration, i.e., the stars in the foreground disk blue plume. These maps are meant to show that our procedure to calibrate the VVV PMs on the astrometric system of Gaia DR3 is correct, and does not introduce spurious patterns across the region analyzed here. }
   \label{Fig:pm_VVV-Gaia}
    \end{figure}

\subsection{Astrometric calibration}

The procedure to derive PMs is described in detail in \citet{contreras+17}. We recall here that VVV, in this region, includes $\sim$100 epochs in the \ks filter and $\sim$5 epochs in each of the J and H filters. 
Our astrometric analysis was performed on all available \ks epochs for each VVV tile. All the RC stars in the NSD region have been measured on at least 70 epochs.  The time baseline covered by these data is of 9 years. The photometry was performed on individual detectors, where stars have a mean FWHM of 0.8 arcsec. The PMs of each star was derived with respect to the $\sim$50 closest RGB stars, chosen as local astrometric reference, within 300 pixels. See \citet{contreras+17} for details.

The initial catalog, therefore, includes relative PMs, i.e., where the mean PM of reference stars is zero, by definition. More recently, however, these PMs were converted to the absolute reference frame, by comparison with the Gaia DR3 catalog \citep{gaiadr3}. Each chip was divided in a 2$\times$2 sub-grid, where we selected well measured, isolated stars in common with Gaia, we measured the median offset between the PMs in both catalogs, and removed it.
In the nuclear region Gaia detects only the brightest stars in optical bands, which are basically all disk stars. This, however, does not constitute a problem, since the difference between the absolute (Gaia) and relative (ours) astrometric system is just a zero point, roughly coincident with the absolute PM of Sagittarius A$^\ast$. In practice, this zero point can be slightly different at different galactic coordinates for at least two reasons.
First, because at large longitudes the net rotation of bulge stars is slighly different from zero. Second, because
occasionally the reference stars are not evenly distributed along the line of sight, and therefore there might be
an excess of stars moving eastward with respect to their counterpart moving westward. This is the reason why we preferred to not trust that relative PMs are truly relative to the GC, and to independently calibrate each 1/4 
chip to the Gaia astrometric system. The number of stars in common with Gaia is $\sim$400 in each 1/4 chip, in the region analyzed here. Very large extinction spots have less stars in common, with the minimum being 80. The uncertainty in the derived zero point is always well below 0.1 mas/yr.

Figure~\ref{Fig:pm_VVV-Gaia} shows a map of the median differences between the VVV final, {\it calibrated} catalog, and Gaia DR3, in the two PM coordinates $\mu_b$ (top) and $\mu_l$ (bottom). Again, these maps are based on the foreground disk stars in common between the two catalogs, not the RC stars analyzed here. However, the astrometric
reference system is independent from the population to which the target star belongs to, therefore if we prove that it is correct for disk stars, then it is correct for bulge RC stars too. Most importantly, these maps prove that the differences we see between the innermost NSD region and its surroundings are not artificially introduced by an offset in the astrometric zero point.

In what follows we analyze the PMs in galactic coordinates, ($\mu_l$ cos$b$, $\mu_b$). In order to shorten the figure labels, we use the expression $\mu_l$, both in the figures and in the text, for consistency, although its value does include the cos$b$ factor.

As explained above, our catalog was constructed on a chip-by-chip basis, and astrometrically calibrated on a 1/4 chip basis. In a final step, we combined all the chips in a single catalog, by using a concave hull to detect the borders of chips, and then of tiles. We removed about 60 pixels from each border of the detector, because of well characterized lower performance, and then we simply kept one of the two sides of the overlapping regions.

\begin{table*}[ht]
      \caption[]{Photometry and Proper Motion catalog.} \label{Tab:cat}
       {\scriptsize
      \begin{tabularx}{\linewidth}{cccccccccrrrr}
            \hline \noalign{\smallskip}
ID             &  long    &    lat   &   J   &  err~J   &   H   &  err~H   &  \ks   &   err~\ks  & $\mu_l$ & err~$\mu_l$ & $\mu_b$ & err~$\mu_b$ \\
               &  (deg)  &  (deg)  & (mag) & (mag) & (mag) & (mag) & (mag) & (mag) & (mas/yr) & (mas/yr) & (mas/yr) & (mas/yr)  \\
            \noalign{\smallskip}
            \hline
    
  b333\_108\_2971 & $-$0.877009 & $-$0.271427 & 19.1735 & 0.094 & 15.812 & 0.047 & 14.333 & 0.006 & $-$8.771 &  3.32 & $-$5.254 &  2.36  \\
  b333\_108\_3885 & $-$0.876231 & $-$0.273262 & 19.7095 & 0.261 & 16.348 & 0.025 & 14.552 & 0.004 & $-$6.785 &  1.72 & $-$1.840 &  1.66  \\
  b333\_108\_7628 & $-$0.876176 & $-$0.280418 & 20.1495 & 0.193 & 17.157 & 0.058 & 15.202 & 0.006 & $-$5.977 &  1.86 & $-$0.575 &  1.85  \\
  b333\_108\_6929 & $-$0.876953 & $-$0.279068 &  --     &  --   & 17.301 & 0.343 & 15.389 & 0.007 &    1.004 &  1.96 &    9.175 &  2.95 \\
 ... & ... & ... & ... & ... & ... & ... & ... & ... & ... & ... & ... &  ... \\ 
           \noalign{\smallskip}
            \hline
         \end{tabularx}
        \tablefoot{The full version of this table is provided in electronic form only.}
    }
   \end{table*}

   \begin{figure}
   \centering
   \includegraphics[width=\hsize]{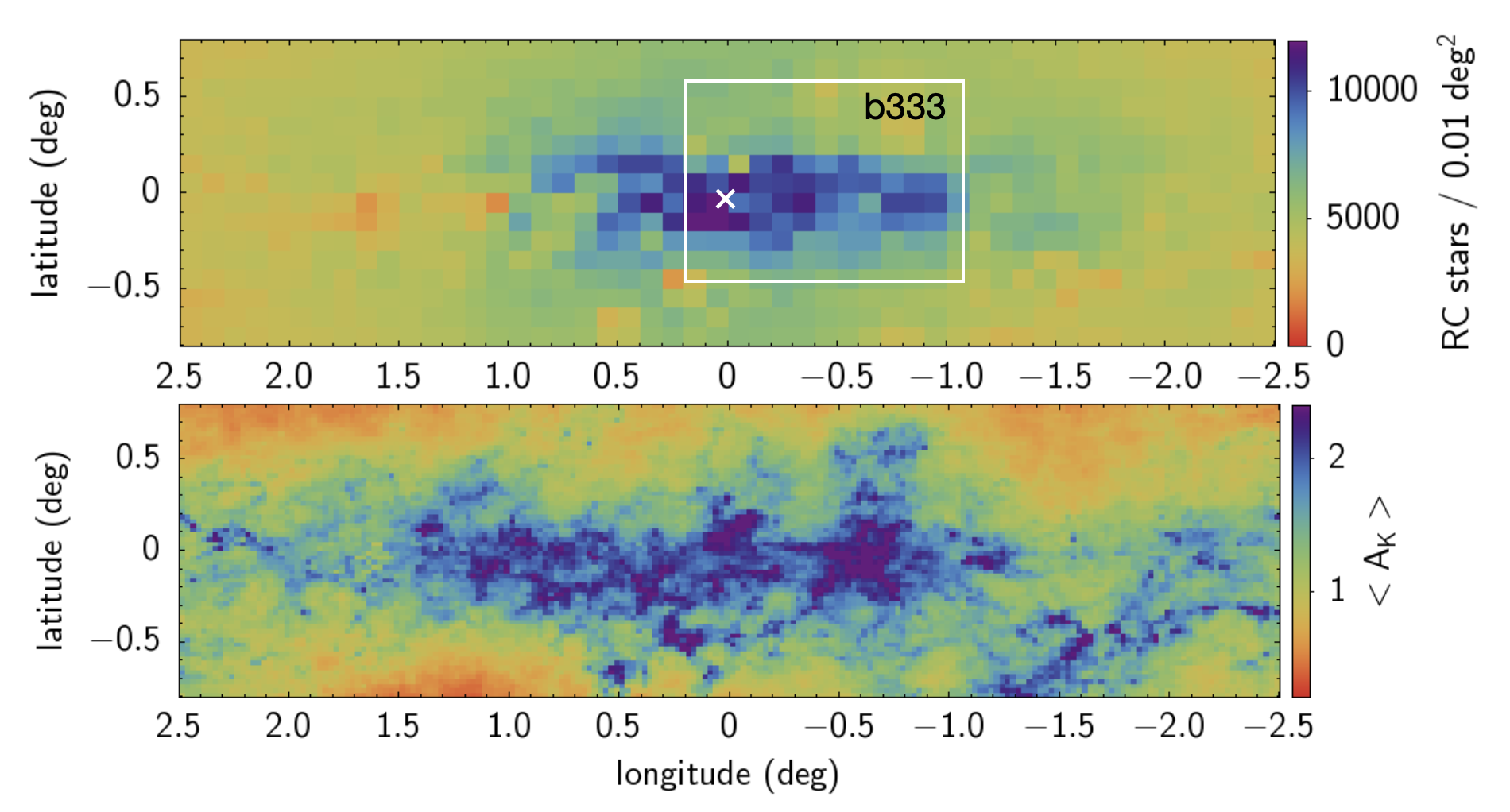}
   \caption{Region analyzed in this work in galactic coordinates. The map in the top panel is color coded by the projected surface density of RC stars, as observed, i.e., without completeness correction. A white cross shows the position of the GC and the coverage of the VVV b333 tile.
   The bottom map is color coded according to the median extinction A$_{\rm Ks}$ (see text for details). }
   \label{Fig:density}
    \end{figure}

\subsection{Spatial coverage and astrometric precision}

This work is focused on the region of the Nuclear bulge including the NSD, and its surroundings, as shown in Fig.~\ref{Fig:density}. The corresponding catalog, whose header is shown in Table~\ref{Tab:cat}, will be 
available in the electronic version of this manuscript. The top panel of Fig.~\ref{Fig:density} shows the projected surface density of RC stars, selected as explained below, without correction for incompleteness. This map, then, does not reflect the real excess of stars in the nuclear region. Indeed, the same map built using the public catalog from the central field of the GALACTICNUCLEUS survey \citep{nogueras-lara+19} reveals that, at $\sim$0.1-0.2 degrees from the GC there are about a factor of 2 more RC stars ($\sim$160/pc$^2$) than in our VVV catalog ($\sim$80/pc$^2$). Nonetheless, we show this {\it observed} stellar surface density map as it shows the number of stars we base our results upon, when we derive the mean PMs of RC stars in this region.
The completeness of VVV at the RC level is already above 80$\%$ at longitude of $l=\pm$1$^\circ$ and a latitude of $b=\pm$0.3$^\circ$.

The lower panel of Fig.\ref{Fig:density} shows an extinction map of this region, based on the excess (H$-$\ks) 
color and the extinction law derived by \citet{nogueras-lara+2021}. Details about the treatment of extinction will
be discussed below, here we just qualitatively show that, as well known, there is a sharp increase in the extinction in the central region. We show the map here in order for the reader to have it handy, and be able to verify whether some of the features observed in the projected kinematics correlates with strong variations in the extinction.

The two maps shown in Fig.~\ref{Fig:density} are intended for illustrative purposes only, in order to show the spatial limits of the present study. Nonetheless, we take the opportunity to make a few comments on their appearance. The region hosting the NSD, i.e., within a radius of $\sim$150 pc ($\sim$1 degree) in the Galactic plane, and of $\sim$45 pc in the vertical direction \citep{launhardt+2002, schonrich+2015}, is certainly different from the sky region close to it, because it shows a sharp increase in the projected stellar surface density (of which Fig.\ref{Fig:density} shows only a lower limit, due to incompleteness), and a similar increase in the interstellar extinction. The latter has a clear physical interpretation: this region hosts the well studied CMZ, a large reservoir of molecular gas that has a major effect in absorbing the light of stars behind it. The excess of stars, instead, although it is certainly there, is not well characterized in terms of both its extension, kinematics, and stellar population. The present work highlights the fact that, precisely because the observational biases such as crowding and extinction are radically different inside this region, it is critical to take them into account when trying to constrain the properties of the NSD. In other words, some observable may show sharp changes inside this region, simply because of the changes in the observing biases.

   \begin{figure}
   \centering
   \includegraphics[width=\hsize]{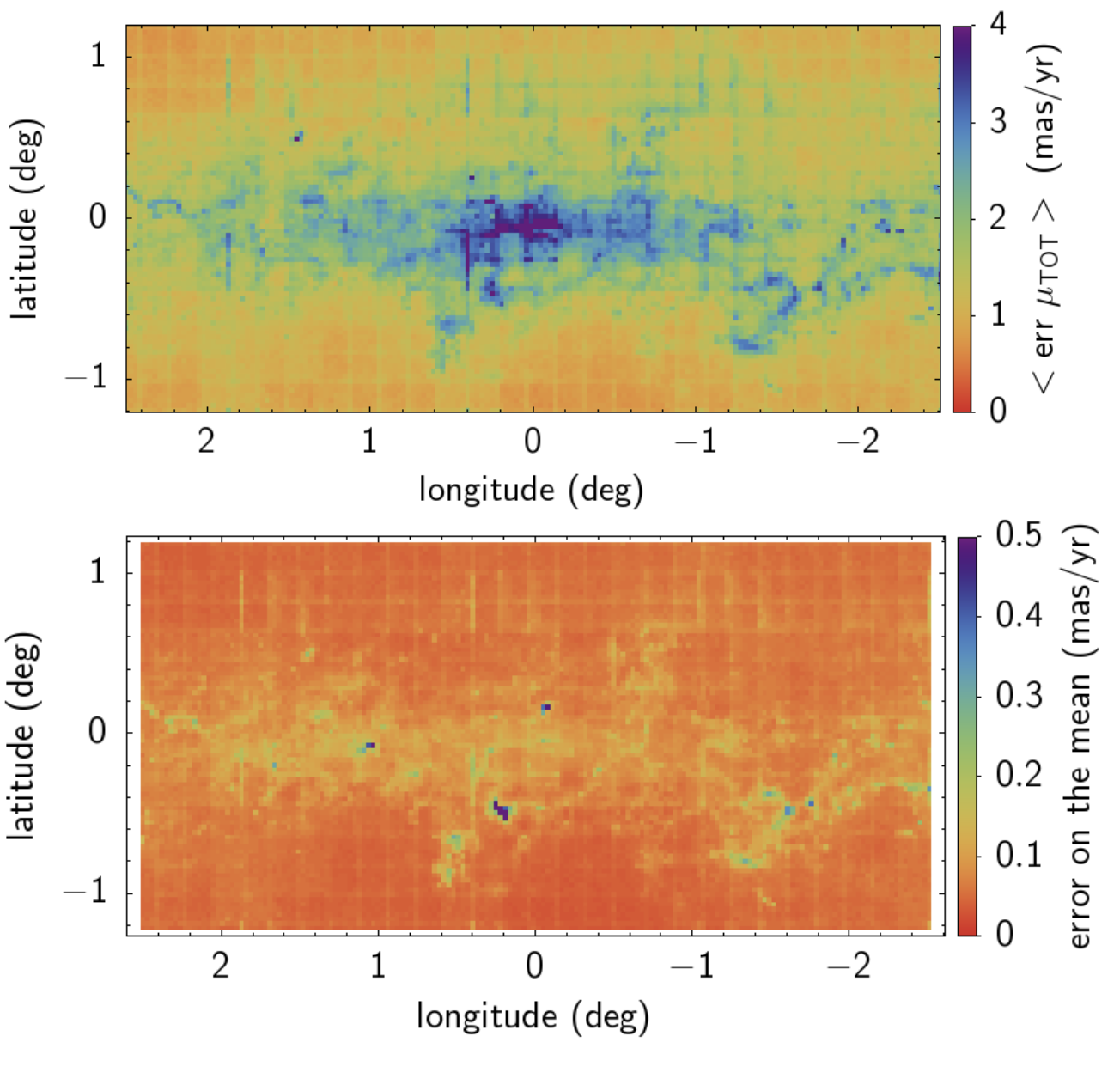}
   \caption{The region analyzed in this work, color coded according to different parameters as shown in the side color bar. {\it Top:} colors indicate the median of the total PM error of individual stars. Because of the combination of crowding and extinction, the error in the individual PM measurements is larger in the NSD region. {\it Bottom:} The
   median PM error of each pixel is divided by the square root of the number of stars (minus one) in each pixel. This map shows the spatial trend of the errors on the mean and median PM, across the sky.}
   \label{Fig:pm_error}
    \end{figure}

Most of the conclusions drawn in the present work rely on the trend of PM for RC stars in the NSD, and the comparison with the surrounding region. For this reason, and keeping in mind the effect that observational biases might have on the results, we show at the top of Fig.~\ref{Fig:pm_error} the map of the median total error in PM, i.e., the median of\\
\[
\sqrt{({\rm err}~\mu_l)^2+({\rm err}~\mu_b)^2}
\]
in each pixel of the map.

We notice that the map shows a squared pattern, due to the fact that the precision in the PM of individual stars becomes worse at the edges of each chip of the VVV mosaic, due to an imperfect correction of spatial distortions. The magnitude of this effect, however, is not very large.

Much more noticeably, the errors on PM are significantly larger in the NSD region, as expected due to the combined effect of crowding and extinction. Nonetheless, because we verified that the errors are symmetric about their mean, 
even stars with relatively large errors contribute to increase the precision of the mean/median, which we use to draw our conclusions. For this reason we did not apply any cut on PM errors. Moreover, although the PM errors are larger in the central region, the number of stars that we measure there are also more than a factor of two larger (see Fig.~\ref{Fig:density}. Because the errors on the mean and median, which we analyze in what follows, are
inversely proportional to $\sqrt(N-1)$, the two effects compensate for each other in such a way that the final error on the mean and median are roughly comparable across the whole area under study, as shown in the bottom panel of Fig.~\ref{Fig:pm_error}. 

\section{Comparison with Shahzamanian et al. (2022)}
\label{sec:shaz}

Before we proceed with the analysis of the PMs of stars in our catalog, we compare it with another recent
determination of PMs in the direction of the NSD, performed by \citet{shahzamanian+22} (hereafter Sh22).
They combined a first epoch HST photometry in the Paschen-$\alpha$ narrow band filter \citep{wang+2010, dong+11}, 
and a second epoch from the GALACTICNUCLEUS program \citep{nogueras-lara+20}, with a time baseline of $\sim$7 years. By analyzing the distribution of longitude PMs, they conclude that it requires three components, two
of them moving toward opposite ways, which they associate with the rotation of the NSD, and a broader one, 
centered at zero, which they associate with the bulge.

Because the spatial resolution of both the first and second epoch photometry of Sh22 is much better than 
that of VVV, it is important to convince the reader that we can provide reliable results with our catalog. To that end, we provide here a direct comparison between the two catalogs. The work by 
\citet{shahzamanian+22} is based on the HST photometry of several noncontiguous fields, covering approximately the region inside $-$0.35$<$$l$$<$0.23 degrees, and $-$0.13$<$$b$$<$0.1 degrees (c.f., their Fig.~2). For this comparison, we selected a rectangular region of our catalog within the limits quoted above.

   \begin{figure}
   \centering
   \includegraphics[width=\hsize]{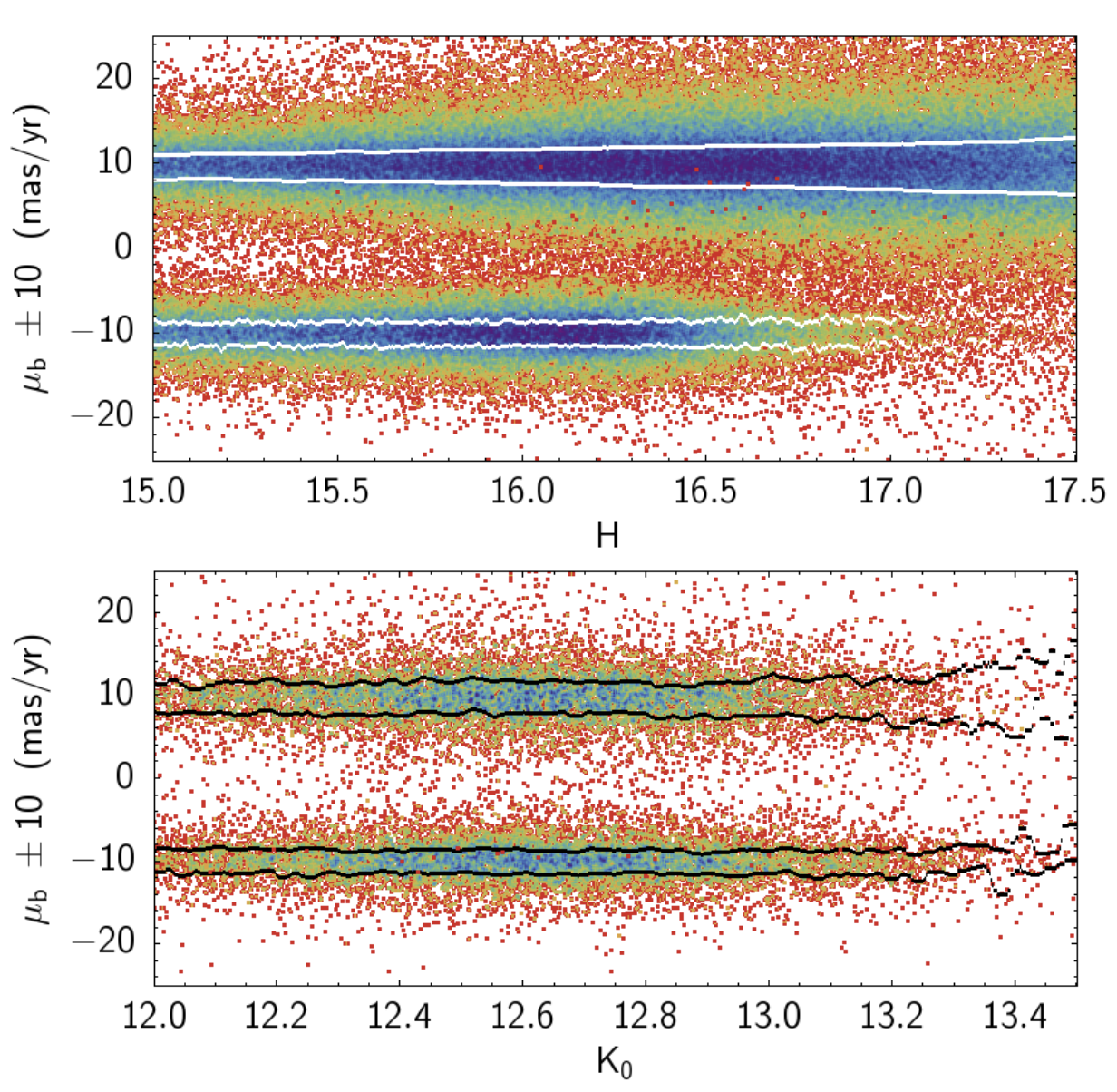}
   \caption{Top panel: Latitude PMs, $\mu_b$ in the VVV catalog presented here (upper distribution) and in the catalog by \citet{shahzamanian+22} (bottom distribution). We added (subtracted) 10 mas/yr to the VVV (Shahzamanian) distribution in order to avoid overlap. The white lines show the IQR for each distribution. Bottom panel: Same as above, but only for the stars in common between the two catalogs, allowing us to use the K$_0$ magnitude in the x-axis. Black lines show the IQR for each distribution.}
   \label{Fig:shaz}
    \end{figure}

The PMs in the two catalogs are very consistent in terms of astrometric calibration. Indeed, the zero point between the present catalog and Sh22 is exactly zero in both (l,b) axis, at magnitude K$_0$=13, i.e., in the middle of the RC. While the difference in $\mu_b$ is flat with magnitude, in $\mu_l$ it has a small slope so that the shift between our $\mu_l$ and Sh22, at K$_0$=12.5, is $-$0.16. We do not make any quantitative analysis for stars brighter than this, and Sh22 quickly run out of stars fainter that K$_0$=13.

In order to assess how the two catalogs compare in terms of depth, completeness, and precision, we plot
the distribution of PM versus magnitude in Fig.~\ref{Fig:shaz}. We make this comparison only for $\mu_b$ because
in this coordinate there is no trend with magnitude (no systematic vertical movement in the MW) and the dispersion 
in PM is directly related to the statistical errors, which is what we want to compare here. 
In the top panel of Fig.~\ref{Fig:shaz} we compare the distribution of $\mu_b$ as a function of H magnitude, in our VVV catalog and in the Sh22 one. 
We plot $\mu_b$ as a function of H magnitude as this is the only magnitude provided in the literature catalog.
The figure shows that the VVV catalog goes deeper than the one by Sh22, and it has larger statistics due to the gaps between fields in the latter (c.f., their Fig.~2). 
Both PMs are centered at zero, but since we are interested in the PM dispersion rather than in the absolute value, we added +10 to our $\mu_b$ and subtracted $-$10 from their $\mu_b$, in order to minimize overlap and be able to
show them in the same panel. As a realistic estimator of the PM error, we use the $\mu_b$ dispersion. Indeed, the observed dispersion is the square root of the PM error, squared, plus the square of the intrinsic PM dispersion of stars in this region. Assuming that the latter is the same, the difference between the observed $\mu_b$ in both catalogs would only depend on the PM error. The white lines in the figure show the interquartile range (IQR) for both distributions. In the common magnitude region, the $\mu_b$ dispersion is smaller in Sh22, but the difference is not dramatic. 

In order to better evaluate the impact of this difference in our analysis, in the bottom panel we show the same thing, this time only for the stars in common between the two catalogs. This allowed us to use the same K$_0$ magnitude in the x-axis. This is important because all the analysis in the following sections will be performed 
using the K$_0$ magnitude of RC stars as a proxy for the distance. 

Black lines show the IQR. At similar magnitude, the standard deviation of the Sh22 PMs is only $\sim$20$\%$ smaller than the one of our VVV catalog ($\sigma$$\mu_b$=2.25 and 2.70, respectively). This is compensated by the fact that our catalog goes significantly deeper, allowing us to estimate the transverse velocity of RC stars further away than 8.2 kpc (down to K$_0$=14), and that it covers a much larger area, allowing us to compare the motion of stars in the region of the NSD, with that of stars in nearby, control regions.

   \begin{figure}[ht]
   \centering
   \includegraphics[width=\hsize]{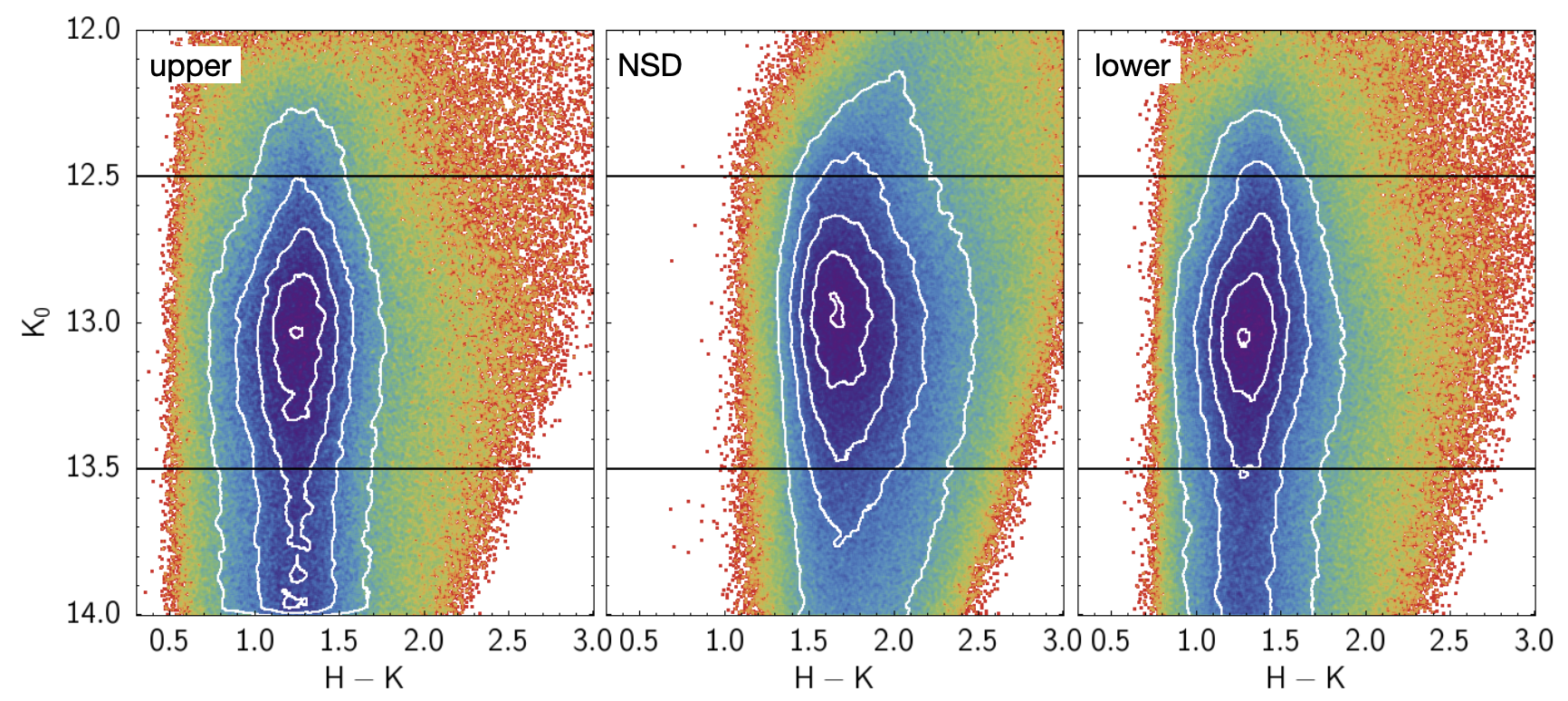}
   \caption{CMD for the selected RC stars in the three regions marked in Fig.~\ref{Fig:pm_maps}. The two horizontal lines mark the magnitude extension of the RC. }
   \label{Fig:cmds}
   \end{figure}

\section{Selection of RC stars}
\label{sec:rc}

It is worth recalling that, when looking at latitude $b$=0 degrees, the line of sight crosses the whole disk and bulge. In order to minimize the contamination from the foreground disk, and to be able to interpret the trend in PM as a function of distance along the line of sight, we restrict the analysis to bulge RC stars. The latter are often used as distance indicators, thanks to the fact that their absolute magnitude does not depend on the age of the stellar population, for ages greater than $\sim$2 Gyr \citep{gallart+05, girardi+16}. Of course, they are not very precise distance indicators, as the RC has an intrinsic spread in magnitude, reflecting both a small luminosity evolution across their lifetime, and the non-negligible metallicity spread of a complex stellar population like the bulge. Nonetheless, because the metallicity is not expected to change dramatically across the region analyzed here, we expect the mean \ks magnitude of RC stars to correlate well with the mean distance of the population, along the line of sight.

In order to automatically select RC stars across the whole region, we start by calculating a de-reddened K$_{\rm s0}$ magnitude (hereafter simply K$_0$), based on the \ejks from \citet{surot20ext}\footnote{The \citet{surot20ext} reddening map is based on the assumption that RC stars in the bulge have a mean intrinsic color of E(J$-$K$_{\rm s}$)=0.635} and adopting the extinction law by \citet{nogueras-lara+2021}, that is, 
\[
{\rm A_{Ks}=0.41\times E(J-K_s). }
\]

The latter was derived for the central region of the GALACTICNUCLEUS survey, which is much smaller than the region studied here. Although it might not be ideal for the whole region, it serves here the purpose to set the mean K$_0$ of the RC approximately constant, so that a selection at 12$<$K$_0$$<$14 encompasses the whole RC everywhere. We also imposed a color selection at (H$-$K$_{\rm s}$)$_0$$>$0.025, in order to exclude foreground main sequence stars. We did not set any upper limit to the color, to avoid introducing undesired biases in the regions with the highest extinction.

\begin{figure}[ht]
   \centering
   \includegraphics[width=\hsize]{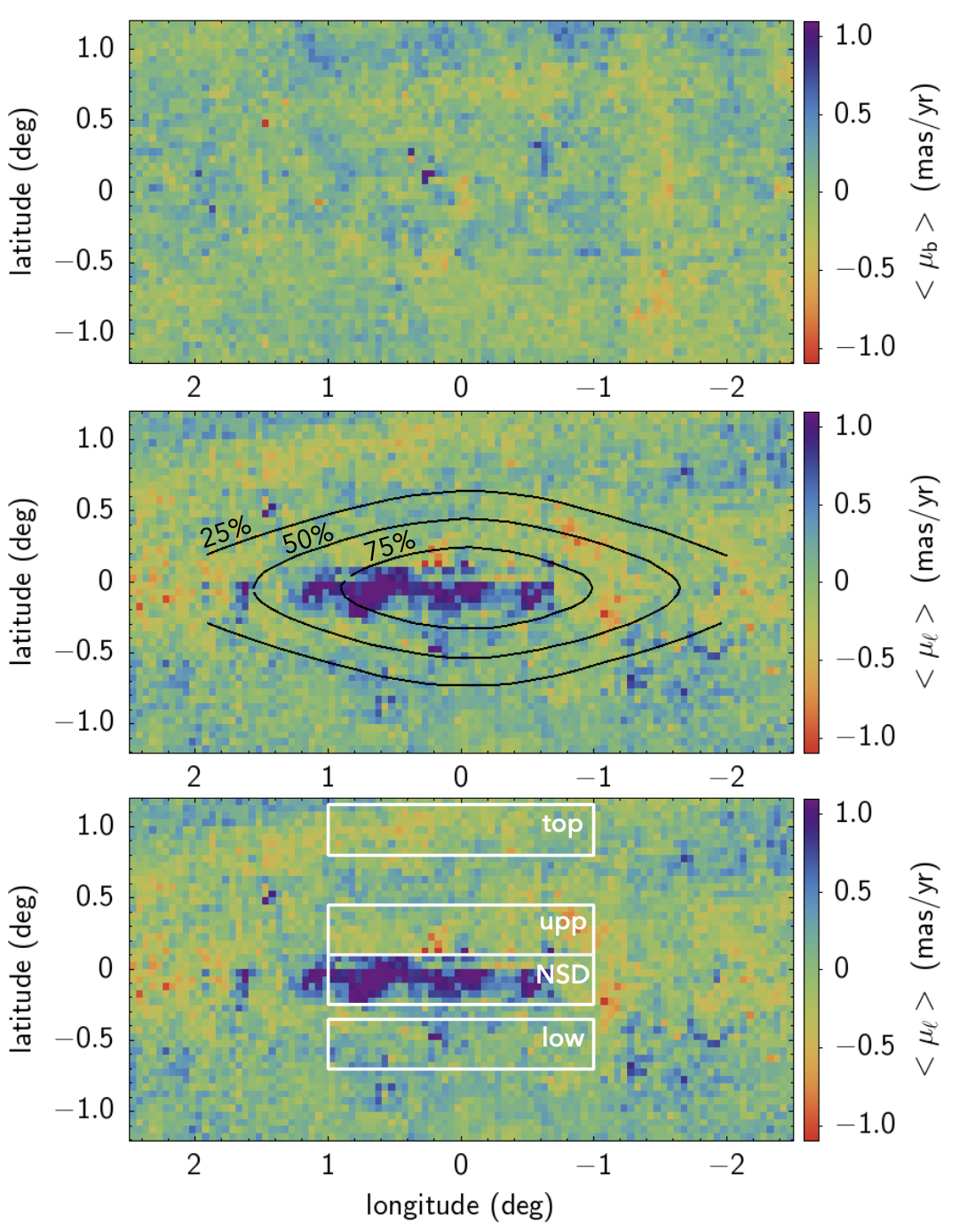}
   \caption{Maps of the median PM in latitude (top) and longitude (middle and bottom) for RC stars in the nuclear region of the MW. As expected, the median latitude PM, $<\mu_b >$ is perfectly smooth and centered to zero, across the whole region. In contrast, the median longitude PM, $<\mu_l >$, shows a clear excess of stars with $\mu_l>0$ in the region of the NSD. In the bottom panel we reproduce the middle panel map, also showing the density contours of the \citet{sormani+22} model (black), and the spatial regions analyzed in the following sections.}
   \label{Fig:pm_maps}
    \end{figure}

Figure~\ref{Fig:cmds} shows the color-magnitude diagrams (CMD) of all the stars within the three rectangles shown in Fig.~\ref{Fig:pm_maps} and labeled as "upper," "NSD," and "lower," respectively. These regions were defined based on Fig.\ref{Fig:pm_maps}, in a way that will be justified below. For the moment let us notice that the peak of the RC is always close to K$_0$$\sim$13, and that our selection was successful at including all the RC stars everywhere, even if the mean extinction is very different in the three regions, as revealed by the different mean color and color spread. 
Two lines at K$_0$=12.5 and K$_0$=13.5 mark the vertical extension of the RC, within which we consider K$_0$ as an approximate distance indicator.

\section{The nuclear stellar disk in proper motion space}
\label{sec:pms}

In this section we examine the mean PM trends, in galactic coordinates, as a function of position in the sky, and along the line of sight. For this and the following analysis, we subtracted from our PMs the motion of Sgr A$^\ast$ ($\mu_l$,$\mu_b$)=($-$6.411,$-$0.219) mas/yr, as derived by \citet{reid+20}, so that in the following figures the motion of the GC will be zero. 

\subsection{Maps of median PM}

Figure~\ref{Fig:pm_maps} shows the map of median PM for all RC stars, in galactic latitude (top panel) and longitude (middle and bottom panel). As expected, the median $\mu_b$ is perfectly smooth and centered to zero across the whole region. This demonstrates that, although the PM errors are larger in the NSD region, this does not affect the median PM.
The median $\mu_l$, instead, shows a clear excess of stars with $\mu_l$$>$0 in the NSD region. Although both the Galactic bulge and the NSD rotate, it is expected that the median $\mu_l$ should be zero because we see both the stars in front of the GC, moving to the east (positive $\mu_l$) and those behind the GC, moving to the west. Even if the NSD is expected to rotate faster \citep{schonrich+2015}, the average between the stars in the near and far side of the Galaxy should be zero. The fact that in the NSD region we see an excess of stars moving eastward suggests that we might be seeing preferentially stars in the near side. This would be reasonable, given that we know that the CMZ is very dense, and therefore it might very well hide an important part of the NSD far side.

Before proceeding to investigate this point further, we define three regions of the sky based on the $\mu_l$ map. A central region, called "NSD" is defined (somewhat artificially) symmetric in longitude, between $-$1$^\circ$$<$$l$$<$1$^\circ$, but slightly shifted toward negative latitudes, at $-$0.25$^\circ$$<$$b$$<$+0.1$^\circ$, in order to capture the observed structure with positive median $\mu_l$. Two comparison regions are defined above and below the "NSD", with the same size. The upper one is at latitude +0.1$^\circ$$<$$b$$<$+0.45$^\circ$, while the lower region has also the same size, but it is slightly shifted at negative latitude in order to avoid what seems to be a fuzzy lower limit of the "NSD" structure. In what follows we use the labels "upper", "NSD" and "lower" to denote these regions. Additionally, we also define a third control region, labeled "top", at +0.80$^\circ$$<$$b$$<$+1.15$^\circ$, outside the limits of the \citet{sormani+22} density model of the NSD. This region will be examined in Sec.~\ref{Sec:long_K0} to double check that the observed absence of a clear kinematical signature of the NSD is not due to the control fields being still inside the NSD itself.

In Fig.\ref{Fig:long_k0} and \ref{Fig:Vrot}, we analyze a larger strip with the same latitude limits but a wider longitude range $-$2.5$^\circ$$<$$l$$<$2.5$^\circ$. This will be obvious from the limits of the x-axis, and/or from the figure labels.

   \begin{figure}
   \centering
   \includegraphics[width=\hsize]{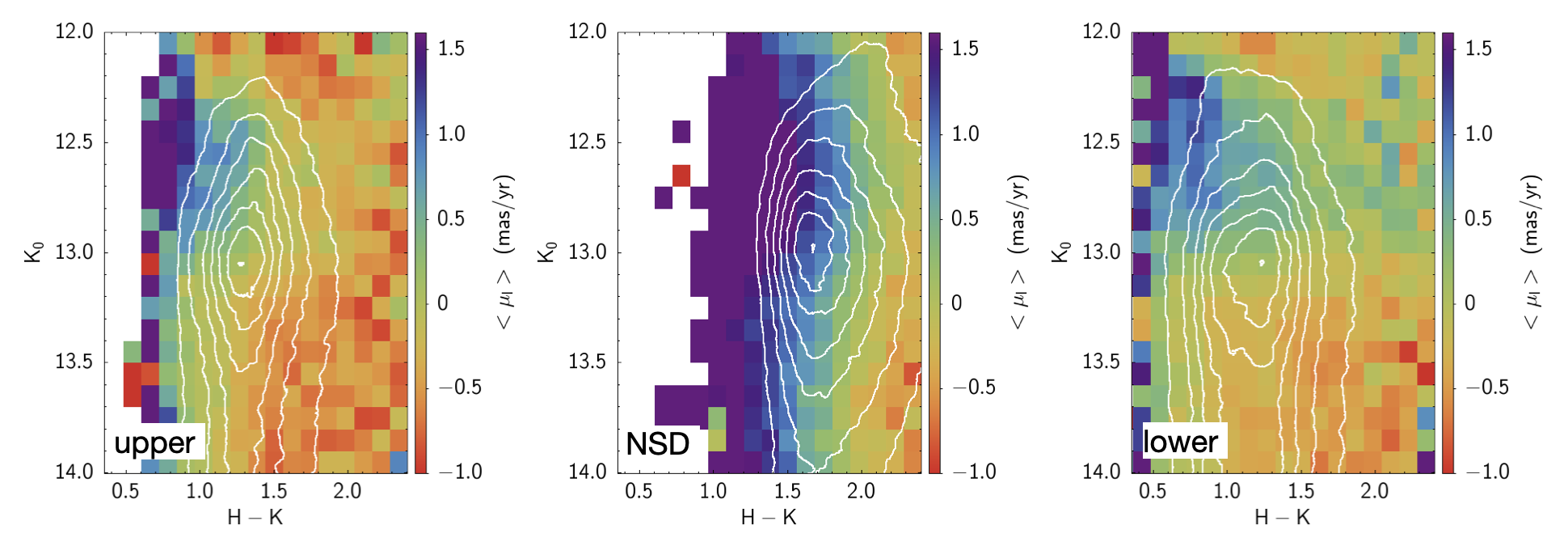}
   \caption{RC region of the CMDs for the three regions defined in Fig.~\ref{Fig:pm_maps}, color-coded according to 2D binned median $\mu_l$ statistics. White contours show the density of stars in the CMD, as in Fig.~\ref{Fig:cmds} to highlight both the peak and the extension of the RC.}
   \label{Fig:cmds_pm}
    \end{figure}

As a first step to confirm the hypothesis that the CMZ is almost completely hiding the far side of the NSD, we plot in Fig.~\ref{Fig:cmds_pm} the CMDs of the three regions defined above, color-coded by median $\mu_l$. The white contours are similar to the ones in Fig.~\ref{Fig:cmds}, showing the density of stars. As expected, in the "upper" and "lower" region we see that RC stars which are brighter than the mean (K$_0$$\sim$13) are blue-ish, i.e., they have a mean positive rotation velocity. RC stars fainter than the mean are, instead, orange, i.e., they have a negative rotation velocity. This trend is much more clear on the blue side of the RC, where reddening is rather low. We recall that the H$-$\ks color is not corrected for reddening in these plots. On the red half of the RC, where stars are more reddened, the above trend seems less clear, but still present. In the "NSD" region, on the contrary, there is no trend of the median $\mu_l$ with K$_0$ magnitude. It is still true that the most reddened stars have a lower mean rotation, possibly because we do see something behind the GCs, but very reddened, and it is harder to locate those stars at the right distance, when reddening is so extreme. All in all, Fig.~\ref{Fig:cmds_pm} confirms the interpretation that we see mainly the near side of the NSD. 

   \begin{figure}
   \centering
   \includegraphics[width=\hsize]{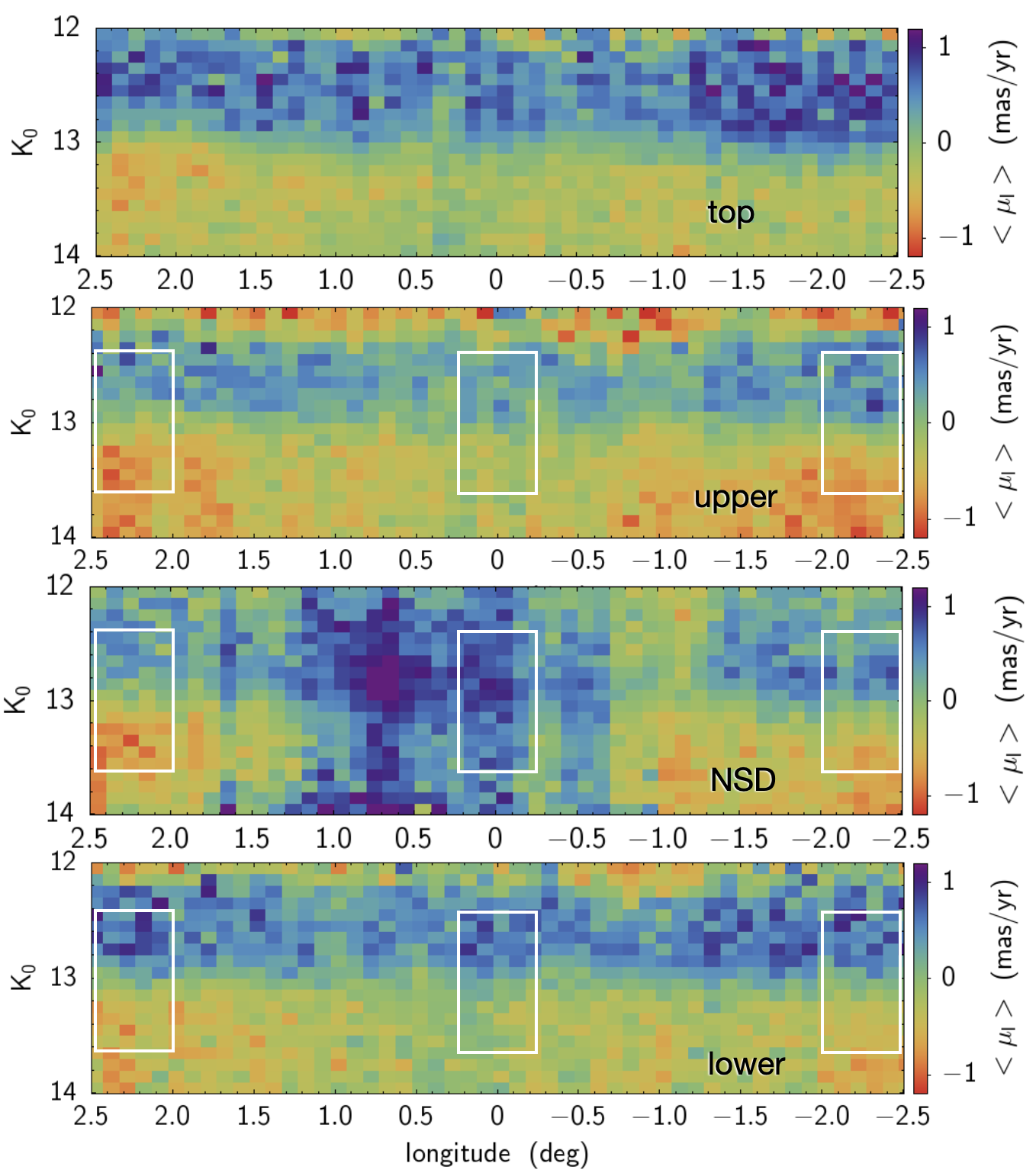}
   \caption{Four panels show data for the four zones defined in Fig.~\ref{Fig:pm_error}, now covering a wider
   longitude range. As indicated by the label, the regions are, from top to bottom: the "top" region, the upper control field, the NSD region, and the lower control field. Within each panel, we show a map of the median $\mu_l$ as a function of longitude in the x-axis and of K$_0$ in the y-axis. Everywhere except in the central region of the middle panel (i.e., toward the NSD) the stars in the upper half of the map show a net positive $\mu_l$, while stars in the lower half show a net negative $\mu_l$. This can be interpreted by assuming that stars in the upper half are brighter than the mean RC, which is roughly at K$_0$$\sim$13, hence closer to us with respect to the GC, while stars in the bottom half are fainter than the mean RC, hence behind the GC. In the central region of the NSD strip, instead, we see only stars with a large positive median $\mu_l$. We interpret this result as an evidence that, in this direction, we see mainly stars in the near side of the NSD. The white rectangles show the regions examined in Fig.~\ref{Fig:Vrot}.}
   \label{Fig:long_k0}
    \end{figure}

\subsection{Mean PMs as a function of magnitude} \label{Sec:long_K0}

As another way of seeing the same thing, in Fig.~\ref{Fig:long_k0} we collapse these CMDs in color, and expand them in longitude. Each panel shows one of the regions of the sky, as indicated by the labels, in K$_0$ versus longitude, again color-coded as a function of median $\mu_l$. In the panel at the top we have added the "top" control region, further away from the NSD, to make sure that there is no significant change in the PM trend in that outer field. In all these panels we see more clearly that, at every longitude and latitude, except in the central region of the "NSD" strip, stars brighter than the mean RC (closer than the GCs) have positive $\mu_l$, while stars fainter than the mean RC have negative $\mu_l$. In the direction of the NSD, instead, all the stars have positive $\mu_l$. In other words, here we are seeing mainly stars in front of the GC. Some of them have K$_0$$>$13, of course, because they are fainter RGB stars belonging to the same population. That is, they are {\it intrinsically} fainter, not more distant.

We hereby consider whether the control fields are far enough, from the NSD region, to allow a comparison free of contamination from the NSD itself. Because the only prominent kinematical structure is the dark blue strip at the center of the $\mu_l$ map, we defined our control fields just outside that strip, above and below, in order to check whether it was the signature of a kinematically distinct component, or due to some selection effect. In addition, it is well known that the bulge rotation curve, even if it has defined as "cylindrical", does show a clear trend with latitude, both in the data \citep[e.g.,][their Fig.~6]{zoc+14_gibs1}, and in the models \citep[e.g.,][their Fig.21]{debattista+17}, with stars closer to the midplane showing faster rotation. Therefore, a proper comparison needs to be done with fields at very close latitude. Nonetheless, according to the recent NSD model by \citet{sormani+22}, reproduced in Fig.~\ref{Fig:pm_maps}, its density decays rather smoothly reaching $\sim$25\% of the total density at |b|=0.5 degrees. Then it would be reasonable to set the control fields further than that. But do we see a kinematical trend that is spatially correlated to the \citet{sormani+22} density model?

   \begin{figure}
   \centering
   \includegraphics[width=\hsize]{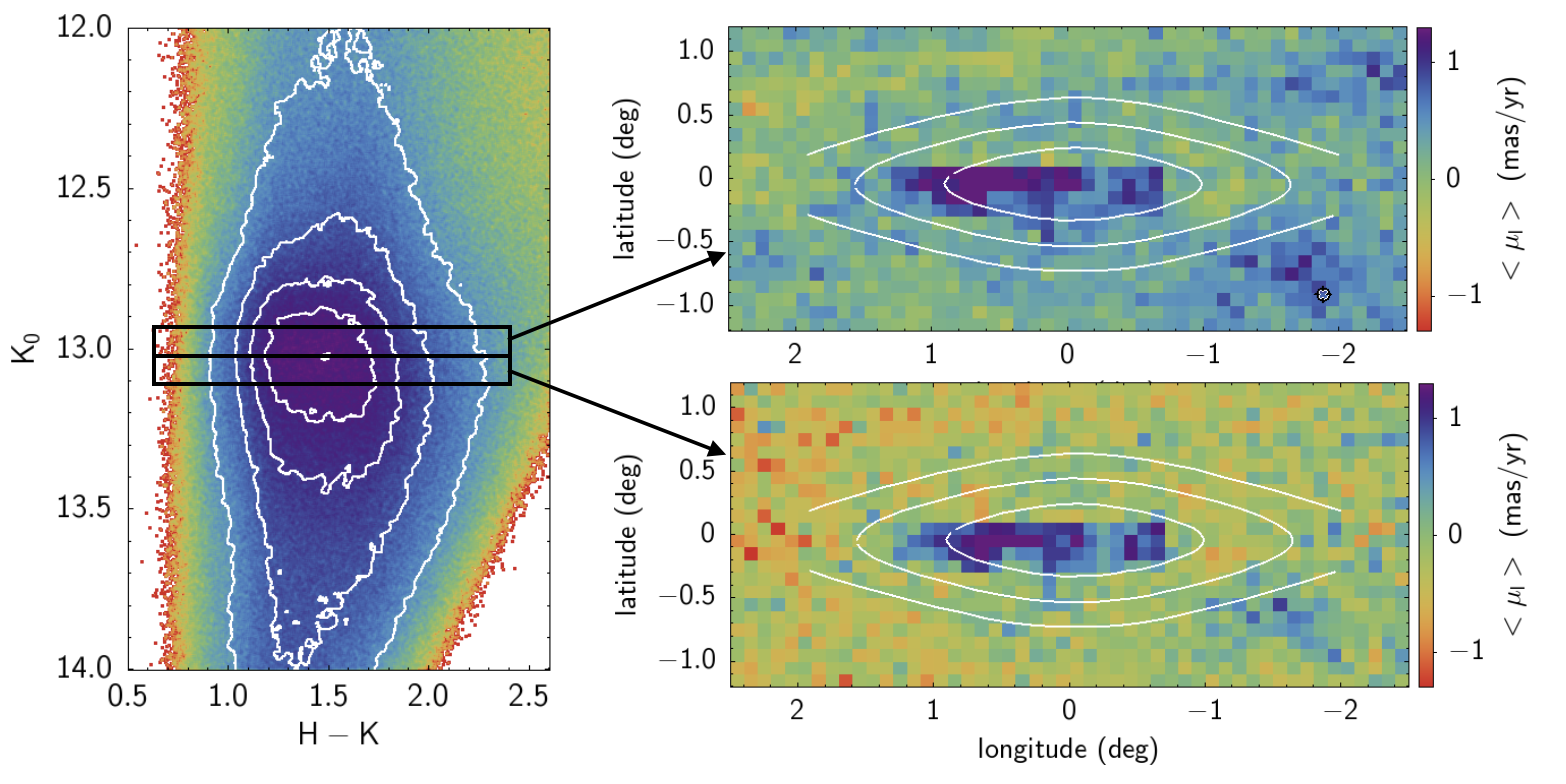}
   \caption{Left: RC region of the CMD for a broad region encompassing the NSD, just for illustrative purposes. Two black boxes highlight stars within $\Delta$K$_0$=$\pm$0.1 mag above and below the RC peak. These two samples should be dominated by RC stars at distances within $\pm$380 pc from the GC (see text for details). Top right: Map of the median longitude PM for the stars $<$0.1 brighter than the RC peak. Bottom right: Same thing for stars $<$0.1 mag fainter than the RC peak. White lines show the density contour of the \citet{sormani+22} model. }
   \label{Fig:kine_sormani}
    \end{figure}

In order to answer that question, we selected only stars $<$0.1 mag brighter than the peak RC, and stars $<$0.1 mag fainter than that, as shown in Fig.~\ref{Fig:kine_sormani} (left panel). The first sample is dominated by stars at distances between 0 and 370 pc closer than the GC, while the second is dominated by stars between 0 and 390 pc farther than the GC. The right panels of the same figure show the maps of the median $\mu_l$ of each of these two samples. As before, we see the dark blue strip that we interpret as due to the higher extinction in the region labeled "NSD", but no other spatial trend correlated with the density contours of the \citet{sormani+22} model. Admittedly, we selected two distance intervals of $\sim$ 380 pc, which are larger than the putative radius of the NSD. We did so in order not to run out of statistics for the calculation of the median $\mu_l$. Still, if this region includes a kinematically distinct component that dominates the density, then we should have seen a strong contamination from it, in the center, not present at the edges. The absence of any trend in these maps reinforces our conclusion that the NSD does not seem to be a kinematically distinct stellar component.

   \begin{figure}
   \centering
   \includegraphics[width=\hsize]{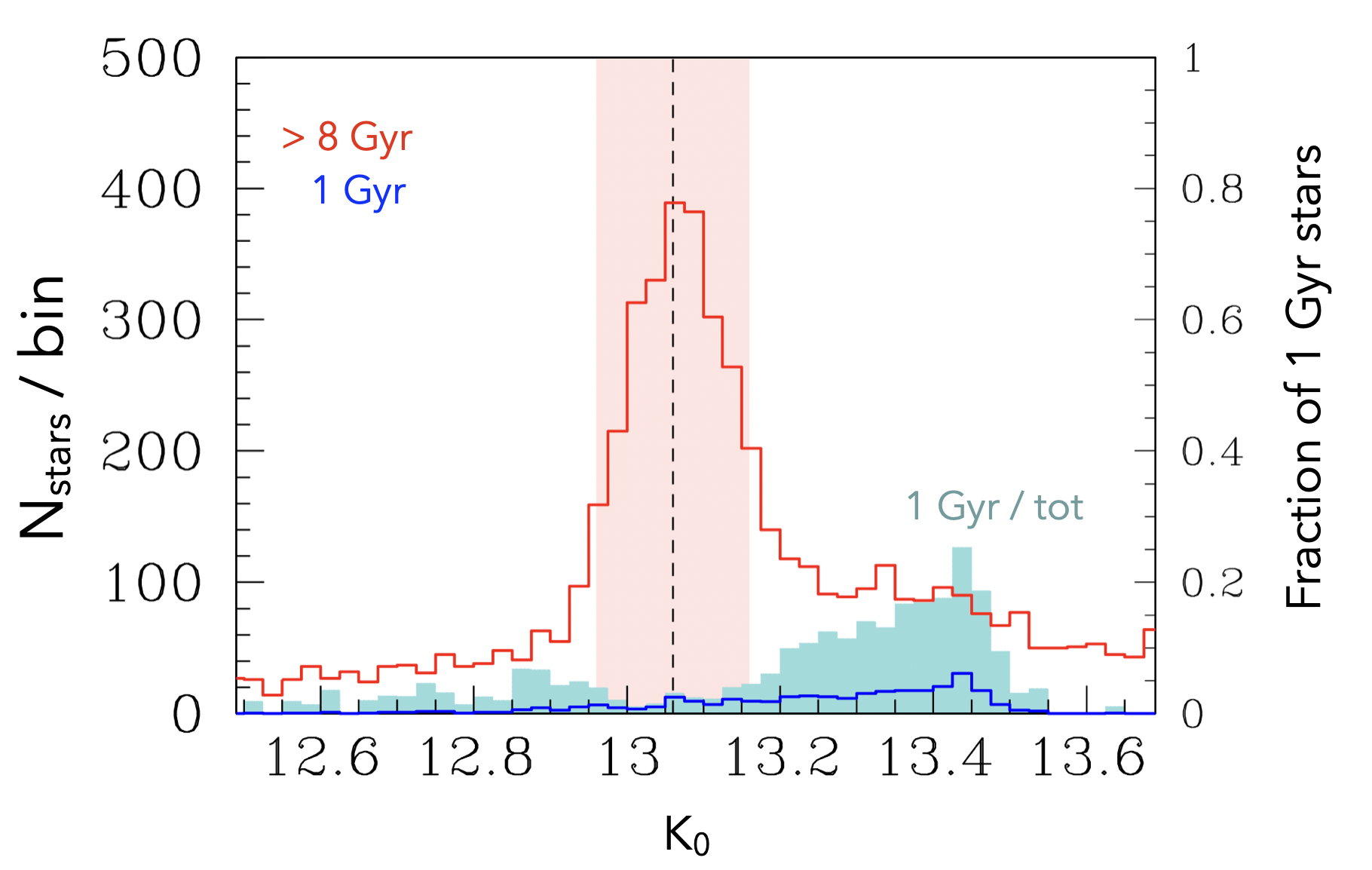}
   \caption{Luminosity function of the RC region for two synthetic populations, both located at 8.2 kpc. The red one has been simulated assuming constant Star Formation History between 8 and 13 Gyr ago, while the blue one has a single burst 1 Gyr ago, making up 5$\%$ of the total stellar mass. The light blue shaded histogram shows the fraction of 1 Gyr old stars, over the total (red+blue), according to the scale on the right y-axis. }
   \label{Fig:LF_age}
    \end{figure}

Addressing a concern by the anonymous referee, we investigated whether the presence of a burst of star formation 1 Gyr ago, making up 5$\%$ of the total mass in stars, {\it only} in the NSD region, as claimed by \citet{nogueras-lara+20}, could affect our results. To this aim, we generated a synthetic population by means of ChronoSynth (see Gallart et al. 2024, in press, for a description of the code), based on solar-scaled stellar evolution models by \citet{hidalgo+18}.
In Fig.~\ref{Fig:LF_age} we show that the LF of a 1 Gyr old population has a RC $\sim$0.4 mag fainter than that of a population older than 8 Gyr. Around K$_0$=13.4, the contamination by this younger population would reach a peak of almost 20$\%$. Nonetheless, the contamination around the RC peak of the old population is just a few percent. Because the analysis in Fig.~\ref{Fig:kine_sormani} is restricted to a region $\pm$0.1 mag around the main RC (shaded pink), the possible presence of a younger population would have negligible effect.  

\subsection{Rotation curves}

   \begin{figure*}
   \centering
   \includegraphics[width=\hsize]{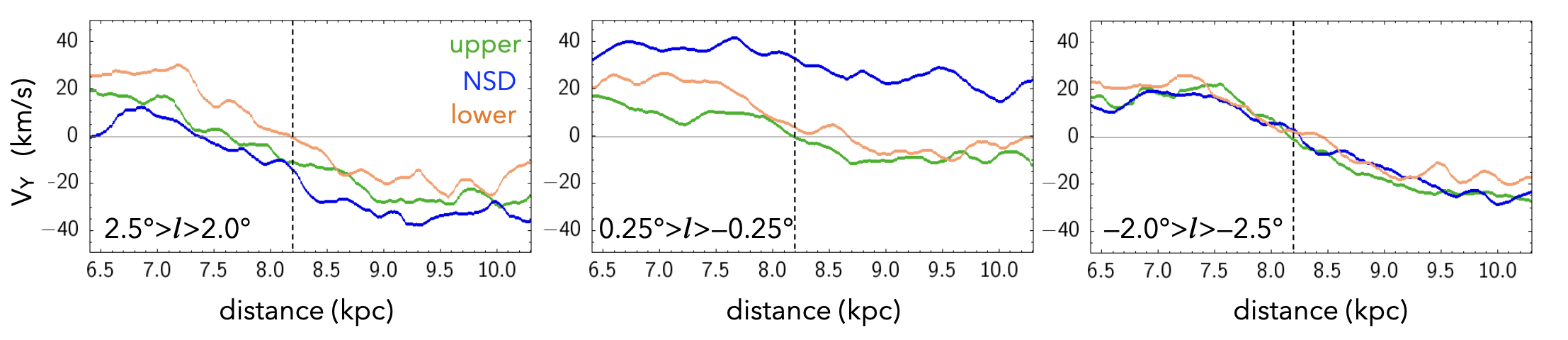}
   \caption{Rotation curves for the nine regions highlighted in Fig.~\ref{Fig:long_k0}, derived as explained in the text. The left panel refers to the regions at positive longitude, the middle one is centered at longitude zero, while the right one refers to the regions at negative longitude. The horizontal gray line is at V$_{\rm Y}$=0 just to guide the eye, while the vertical dashed line marks the distance to the GC. As expected, most rotation curves cross the V$_{\rm Y}$=0 line at the GC position, confirming that the adopted extinction law is fairly appropriate for this region of the Galaxy. }
   \label{Fig:Vrot}
    \end{figure*}

To go one step further with this interpretation, we can actually use the RC as a distance indicator, and convert the K$_0$ magnitude of RC stars into distance. This works only for RC stars, of course, i.e., in the range 12.5$<$K$_0$$<$13.5. In practice, we proceeded as follows. We verified that the mean K$_0$ magnitude of the RC, in Baade's Window, is K$_0$=13.06. 
For Baade's Window, we adopted the extinction law by \citet{nishiyama+09}, appropriate for $|b|$=-1$^\circ$$-$4$^\circ$ \citep{gonzalez+12}. The mean distance of RC stars in Baade's Window is d=8.2 kpc, hence their distance modulus is  
14.57, and their absolute \ks magnitude is M$_K^{\rm RC}$=K$_0^{\rm RC}$$-$14.57=$-1.51$. It is therefore straightforward to convert the observed K$_0$ magnitude of each RC stars in the nuclear region into a distance, using this absolute magnitude and the definition of distance modulus.  The derivation of a distance to each RC star also allows us to convert the observed mean $\mu_l$ into a rotation velocity around the GC. Any vertical cut of Fig.~\ref{Fig:long_k0}, showing $\mu_l$ {\it vs} K$_0$, can be converted into a rotation curve V$_{\rm Y}$ {\it vs} distance. Note that \citet{nogueras-lara+20} concluded that most of the NSD stars were formed at least 
8 Gyr ago, like in Baade's Window \citep{renzini+18, bernard+18}, while 5$\%$ of them were formed $\sim$1 Gyr ago. By means of a simulated luminosity function, we have verified, in the previous section, that the recent burst would not affect the mean magnitude of the RC.
If the star formation history of the NSD region is completely different from both the one in Baade's Window, and
the one derived by \citet{nogueras-lara+20}, then this might affect the mean magnitude of the RC, and the calibration of magnitude into distance would not be correct. Still, this would imply a horizontal shift in 
the rotation curves, which would not affect our conclusions.

We did this exercise for three longitudes, shown by the white rectangles in Fig.~\ref{Fig:long_k0}. The result is plotted in Fig.~\ref{Fig:Vrot}. Each panel here refers to a different (small) longitude range, while curves of different colors refer to the three latitude strips: "upper", "NSD" and "lower". In every direction we see a normal S-shaped rotation curve, except in the direction of the NSD, where the rotation velocity stays positive and high at every distance. Of course, this cannot be true: what is happening here is that stars fainter than the mean RC are simply stars belonging to the RGB below the RC. We are converting their observed magnitude into distance assuming that they are RC stars, which they are not.  Again, we are seeing the complete RC+RGB of stars in front of the GCs, and virtually nothing behind it.

A few considerations are mandatory at this point. 
First, at a distance of 1 kpc from the GC, we are certainly outside the NSD, and the bulge rotation velocity should be closer to 70-80 km/s \citep[e.g.,][]{zoc+14_gibs1, ara+20} rather than 20 km/s as shown in Fig.~\ref{Fig:Vrot}. This is due to the fact that, each K$_0$ magnitude bin is {\it dominated} by RC stars at a given distance, but it also includes RGB stars at different distances. Therefore, the line of sight rotation curve derived here is smoother than the ones derived via RVs of individual stars, as a function of longitude. In fact, this rotation curve is very similar to the ones derived by \citet{clarkson+18} by means of PMs, with an approach similar to the present one. 

In virtue of the same argument, the fact that the NSD rotation velocity is everywhere large does not necessarily means that the NSD rotates faster. In fact, if at the NSD we are seeing only stars in front of the GC, then it is expected that their measured velocity is higher, because we see here a clean sample of stars rotating eastward. In other directions, instead, our sample of RC stars is always contaminated by RGB stars at different distances, rotating in the opposite direction.

A legitimate question at this point is: do the present data prove that the NSD really exist? In order to answer this question we should first define what is the NSD. What would be the unambiguous signature of a NSD? At the moment
there is a clear evidence for the presence of a rather flat structure denser than its surroundings. There is also
evidence of the presence of young massive stars (and clusters) and of dense molecular clouds with active star
formation. However, we find here no signature that this structure is rotating faster than its surroundings. Previous PM studies in lines of sight toward the NSD, such as \citet{martinez-arranz+22, shahzamanian+22, nogueraslara+22los} report the detection of the NSD rotation from the evidence that the $\mu_l$ distribution requires three Gaussians to be fitted. Although \citet{martinez-arranz+22} provide evidence that the most reddened stars rotate in the opposite direction of the least reddened ones (a fact that we also detect) they provide no comparison with nearby fields. Therefore, they cannot state that the NSD rotates faster than the MW bulge at that position.
In fact, we see the same trend of $\mu_l$ with color (= extinction) also in the comparison fields above and below the NSD. The present study provides, for the first time, a curve of rotation velocity versus distance, in a line of sight toward the NSD, {\it as compared to other lines of sight above and below it}.

The robust conclusion of the present work is that the main difference between what we see at the NSD and what we see elsewhere is that the CMZ hides almost completely the NSD far side. In any other nearby direction, instead, we see stars both in the near and in the far side, rotating in opposite directions. As expected in these circumstances, the mean $\mu_l$ of RC stars at the NSD is larger than in any other direction, as it is not contaminated by more distant upper RGB stars with negative $\mu_l$.

\section{The nuclear stellar disk in radial velocity space}
\label{sec:RVs}

   \begin{figure}
   \centering
   \includegraphics[width=\hsize]{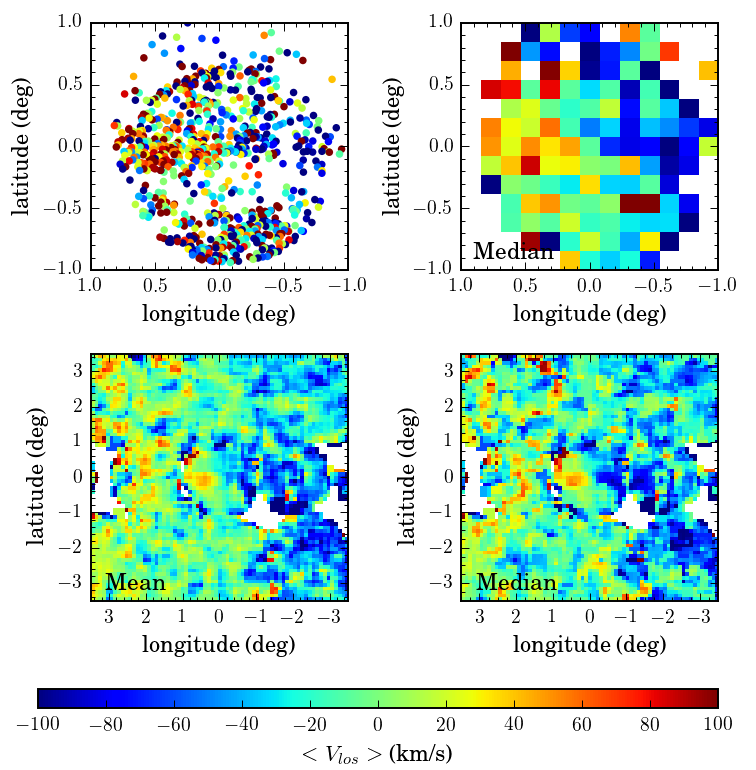}
   \caption{Maps of the line of sight velocity (V$_{\rm los}$) of bulge stars in the APOGEE DR17 data. \textit{Top left:} the central plate of the APOGEE survey (called GALCEN) with individual target stars color coded according to their V$_{\rm los}$. \textit{Top right:} same data binned. Colors show the median V$_{\rm los}$ in each coordinate bin. 
   \textit{Bottom:} map of mean (left) and median (right) V$_{\rm los}$ from an APOGEE DR17 sample covering a larger spatial area. Each bin contains the statistic computed from stars located over a region 3.5 times larger than the pixel itself. The color bar at the bottom applies to all the panels.}
   \label{Fig:NSD_RV}
    \end{figure}

The first kinematical detection of the NSD is reported by \citet{schonrich+2015} who analyzed all the stars observed in the central field (the GALCEN plate) of the APOGEE survey, detecting a flat structure rotating much faster than its surroundings, with peak velocity at $\pm$120 km/s . From their Fig.~1, the peak velocities are seen at approximately longitudes of $\pm$0.6 degrees, corresponding to a radius of $\sim$100 pc, although they quote R$_{\rm NSD}=150$ pc.  Upon careful inspection of their Fig.~1, it appears that they optimized the color scale in such a way to hide the bulge rotation velocity, and highlight the faster velocity of the NSD. In fact, all the values between V$_{\rm Y}$=+30 km/s and V$_{\rm Y}$=$-$50 km/s are plotted in the same tone of green, and as a consequence the bulge rotation curve is completely flat from $l$=$-$3$^\circ$ to $l$=+4$^\circ$. 

We attempted at reproducing this result from \citet{schonrich+2015} with the latest data release of APOGEE, DR17\footnote{The data were retrieved from the SDSS-IV archive at \url{https://www.sdss4.org/dr17/irspec/spectro_data/.}} \citep{apogee, apogeeDR17}. We anticipate that the signal of the presence of the NSD is not as clear as it seemed in the previous analysis.

The top left panel of Fig.~\ref{Fig:NSD_RV}  shows the sky distribution of the GALCEN sample, as available in the \texttt{allStar}\footnote{\url{https://data.sdss.org/sas/dr17/apogee/spectro/aspcap/dr17/synspec_rev1/allStar-dr17-synspec_rev1.fits}} file provided by the collaboration, color coded by their line-of-sight velocity (V$_{\rm los}$).

Both the GALCEN plate, and its surroundings now include many more targets. Following \citet{schonrich+2015}, and after removing telluric standards (by their APOGEE TARGFLAGS), we tried to reject stars with low extinction, in order to remove foreground stars. We encountered two problems with this. First, by applying the same selection as \citet{schonrich+2015}, i.e., rejecting stars with A$_{\rm Ks}$$<$3, we were left with no stars in the whole plate. Second, we noticed that, in the NSD region, many stars do not have the AK$\_$TARG extinction value provided by the APOGEE collaboration from the RJCE method \citep{Majewski2011}. Applying a selection on A$_{\rm Ks}$, then, would introduce a bias that is different in the central region with respect to the rest. 
On the other hand, the target selection for the GALCEN plate was intended to provide a clean sample of stars located in the central inner Galaxy \citep{zasowski+13, zasowski+17, schultheis+20}. Of course, some foreground stars might remain, but its fraction is expected to be low. Therefore, because the density of the small number of foreground stars should not vary dramatically toward $b$=0$^\circ$ compared to $b$=1$^\circ$-2$^\circ$, and in order to avoid introducing a different bias for the NSD region with respect to the surrounding bulge, we do not apply any selection based on A$_{\rm Ks}$

We can see in the top left panel of Fig.~\ref{Fig:NSD_RV} a hint of a larger concentration of stars with positive/negative  velocities at positive/negative longitudes. To verify if this corresponds to a clear spatial signal, the top right panel shows a median map of V$_{\rm los}$. We do not retrieve any significant spatial trend of V$_{\rm los}$, with only a slight overall asymmetry of the pixel values around the minor axis.

In the bottom panels of Fig.~\ref{Fig:NSD_RV} we attempted a different approach. In this broader visualization of GALCEN and the surrounding bulge field population, the sample was divided into a regular array of bins. Each bin represents the mean/median (left and right panels, respectively) V$_{\rm los}$ of all stars across a region 3.5 times larger than the size of the bin in the visualization. This strategy attempts to increase number statistics, in order to detect any signal concerning the overall motion of stars at both sides of the minor axis. By doing this, we assume that if such a signal exists, we would stabilize it against low number statistics in the spatial region where it dominates, but at the cost of blurring its borders. The two bottom panels serve to visualize the effect of using two different statistics over the pixels. In both cases, we can see an overall large scale trend of the stars having positive net V$_{\rm los}$ at positive longitudes, and negative V$_{\rm los}$ at negative longitudes, revealing the bulge rotation. Overlapping this signal, there is a degree of stochasticity in the form of small scale enhancements of positive and negative velocities. However, a qualitatively larger signal of bulk motion can be seen at the putative NSD location. Indeed, there is a hint of a bipolar distribution of V$_{\rm los}$, with maxima at about $l=\pm0.5^\circ$. We emphasize that, while this is a qualitative detection of the NSD in radial velocity, this only shows the plausibility of the presence of such a signal. With the current size of the sample, we cannot fully confirm its presence, nor provide a detailed account of the spatial and kinematic properties of the NSD.

A rotation curve for the NSD was derived by \citet{schultheis+21} by means of KMOS data. In their Fig.~10 they show that the measured rotation velocity at the NSD is higher than the bulge velocity derived from APOGEE data, at 0.3$^\circ$$<$$b$$<$4$^\circ$. In fact, because the bulge rotation increases with decreasing latitude, the comparison should be done either with stars at very similar latitudes, or with a proper extrapolation of what the rotation should be, at b=0 deg, in the absence of a NSD. 

The forthcoming near infrared fiber spectrograph MOONS at the ESO Very Large Telescope \citep{ciras+20-MOONS} will most likely yield the final answer on this issue.
  
\section{The physical properties of the NSD}

Can we use the present data to constrain the structural and kinematical properties of the NSD?
We have already discussed the fact that, due to the contamination by RGB stars at different distances, the line-of-sight rotation curve obtained from the PM of RC stars yields peak velocities lower than the real ones. Therefore, the highest velocity observed at the NSD region, compared with its surroundings, cannot be easily interpreted in terms of the NSD rotating faster. Rather, it is another indication of a lower contamination from stars at the far side, rotating in the opposite direction.

\begin{figure}
   \centering
   \includegraphics[width=\hsize]{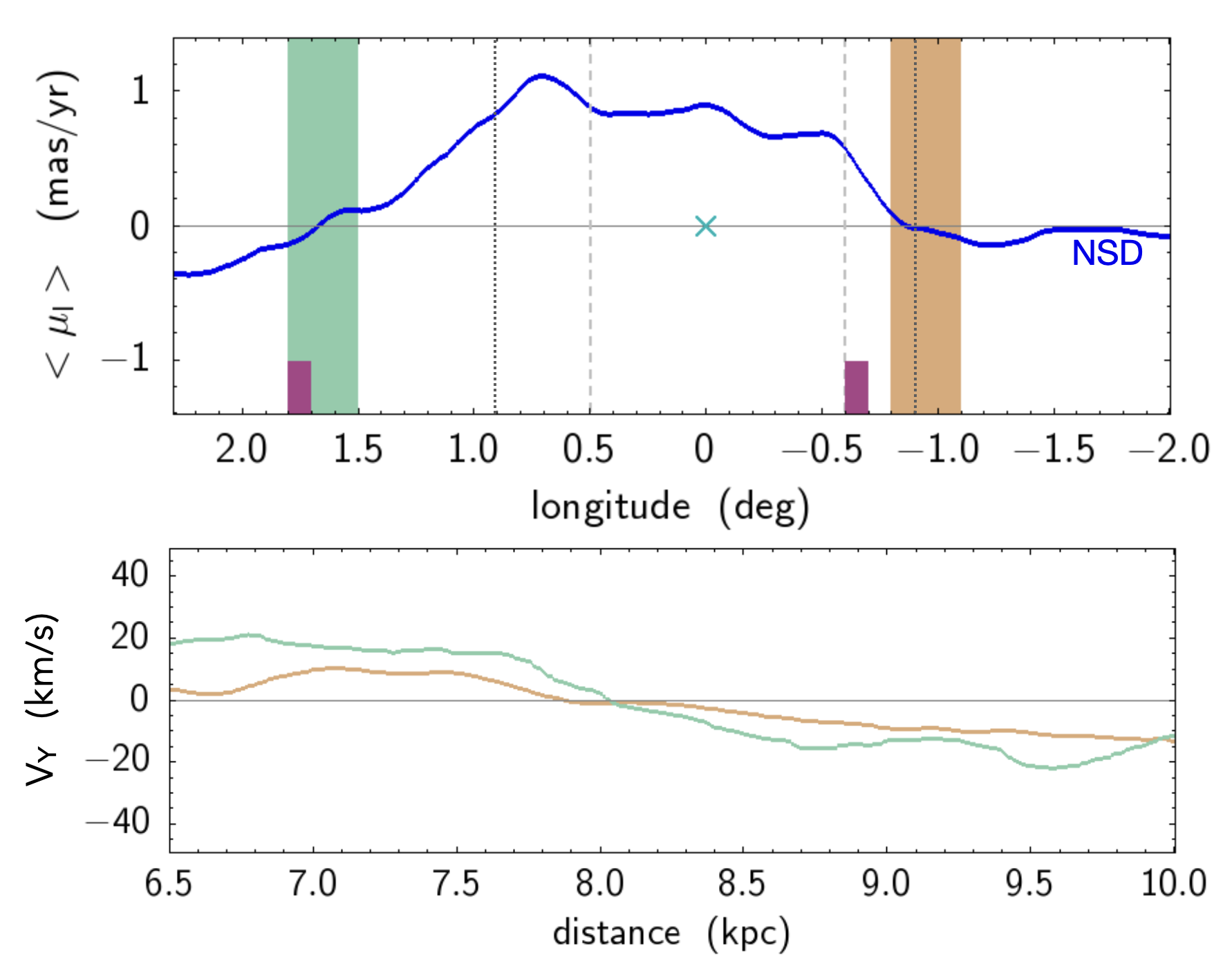}
   \caption{\textit{Top}: This panel shows the running median of the longitude PM, $\mu_l$, for RC stars in the range K$_0$=12.5-13.5, as a function of longitude, for the NSD strip. The green and brown rectangles mark the approximate locations where the longitude velocity goes to zero, as possible edges of the NSD. The purple rectangles mark the approximate edges of the CMZ. Gray dashed lines are the positions at which \citet{schonrich+2015} found the RV positive and negative peak, respectively. The cross marks the position of the GC, while a thin black line shows V$_{\rm Y}$=0. \textit{Bottom}: Rotation curves corresponding to the green and brown rectangles of the top panel. See text for details.} 
   \label{Fig:Vrot_edges}
    \end{figure}

In principle, the shape of the rotation curve could be used to try and constrain the extension of the NSD.
The top panel of Fig.~\ref{Fig:Vrot_edges} shows the trend of median $\mu_l$ as a function of longitude, for NSD strip. As clear from the map in Fig.~\ref{Fig:pm_maps}, of which this curve is a horizontal section 0.35$^\circ$ wide in latitude, the median $\mu_l$ is roughly zero at large longitudes on both sides, while it is positive close to the $l$=0 direction. If we take this as a feature of the NSD, then we could use the two edges where this curve goes to zero, as the physical limits of the NSD. Indeed, at the physical edges of the NSD, there must be two points in which the rotation vector is parallel to the line of sight, and therefore all the velocity is radial, with the component along the PM being exactly zero. We mark these two edges with the green and brown rectangles. These positions should coincide with the RV peaks found by \citet{schonrich+2015}. From their Fig.~1, these peaks occur at $l$=+0.5$^\circ$ and $l$=$-$0.6$^\circ$, respectively, and these positions are shown as gray dashed lines here. Clearly, they are much closer to the GC than the tentative edges we detect here. Figure~\ref{Fig:Vrot_edges} also shows the peaks found by \citet{schultheis+21}, as vertical lines at $l$=$\pm$0.9. These measurements also suggest a smaller NSD, although the one at negative longitude coincides with the tentative edge detected here.

It is interesting to note that the tentative edges of the NSD detected here are not symmetric about the GC. Indeed, the CMZ is also strongly asymmetric \citep{bally+88}, and it makes sense that, if we interpret the large positive $\mu_l$ values as an absence of the opposite negative $\mu_l$ values hidden by the CMZ, then this excess should coincide with the size of the CMZ. The latter is marked by the purple large ticks. At negative longitude this is the border of the Sgr C cloud, while at positive longitude this is the edge of the group associated with the 1.6$^\circ$ cloud \citep[see, e.g., Fig~1 in][]{battersby+2020}. These positions are qualitatively consistent with the edges detected here, confirming our hypothesis that we see this large positive $\mu_l$ because of the presence of the CMZ, hiding the far side of the NSD.

If the NSD is a separate structure rotating faster than its surroundings, however, we would expect that at the edges detected above, not only the mean $\mu_l$ should be zero, but also the line of sight rotation curve should be flat. At the edges of the NSD, in fact, we would expect that stars at 8.2 kpc, but also slightly closer, should all have a rotation velocity with a null longitude component. The bottom panel of Fig.~\ref{Fig:Vrot_edges} shows the rotation curve at each of the two edges of the NSD.
Again, the result is not clear. While the brown curve shows a flattening in the central part, the green curve is very similar to all the other rotation curves in Fig.~\ref{Fig:Vrot}.

In summary, the present data do not allow us to establish whether the NSD is a separate structure with a different rotation. Our data are consistent with the hypothesis that we are seeing a different kinematic patterns due to the sole effect of the CMZ hiding the back side of this region. The fact that APOGEE DR17 data also do not confirm the clear detection by \citet{schonrich+2015} highlights our poor understanding of the extension, kinematics, and even the existence of a cold, fast rotating NSD at the center of the MW.

\section{Summary}

We have presented the kinematics of the NSD region by means of PM measurements obtained from PSF photometry on VVV survey data.
The analysis was restricted to RC stars in order to be able to relate their mean dereddened magnitude, K$_0$, with their mean distance. While the typical PM error in the innermost region is $2-3$ mas/yr, the present analysis
is based on the average PM of a large number of stars, therefore improving the precision by a factor of sqrt(N).
For instance, Fig.~\ref{Fig:long_k0} is based on $\sim$400 RC stars, in each bin of $\Delta$K=0.1 mag and $\Delta$$l$=0.1 degrees, resulting in an error on the mean $<\mu_l>$=0.10-0.15 mas/yr, which is significantly smaller than the
systematic variations of $\gtrsim$1 mas/yr, that is consistently seen across several pixels of those maps.

The analysis of the median longitude PM, $\mu_l$, across the plane of the sky and as a function of the de-reddened K$_0$ magnitude 
showed that, as expected due to the bulge rotation, the stars that are on the near side of the bulge have a net positive $\mu_l$ (rotate eastward), while the stars at the far side have a net negative $\mu_l$ (rotate westward). A map of the median $\mu_l$ across the plane of the sky, therefore, averages to zero PM. When looking at $-$1$<$$l$$<$1, however, we see only stars with a positive $\mu_l$. We interpret this as a evidence that, in this direction, we see only stars at the near side of the NSD. This hypothesis is consistent with our understanding of the NSD as a ring of stars with a large stellar density enclosing the CMZ. The dense molecular clouds constituting the latter would hide the far side of the NSD.

By converting de-reddened K$_0$ magnitudes for RC stars into distances and PMs into space velocities, we obtained rotation curves along different lines of sight in the MW nuclear region. In the direction of the NSD, the curve is rather flat at $\mu_l$=30 km/s; whereas in nearby regions outside the NSD, the curve has the usual S shape from positive to negative velocities, crossing zero at 8,2 kpc, as expected. 
The peak rotation is lower than that observed with RVs at different longitudes. This is not surprising given that RC stars at a given distance are contaminated by RGB stars closer and further away. The derived rotation curve, however, is consistent with other measurements obtained in the bulge with the same method as \citet{clarkson+18}. \\

The main results of the present paper are the following:
\begin{itemize}
\item We detect the rotation of the nuclear region of the MW. The brightest half of the RC does indeed move eastward, while the faintest half moves westward, in longitude. \smallskip

\item We detect a structure, confined within longitude $-$1.5$<$$l$$<$0.8 and latitude $-$0.25$<$$b$$<$+0.1, where all the RC stars move eastward.  We interpret these observations as evidence that the CMZ hides the far side of this structure.  \smallskip

\item We do not find clear evidence, in the present PM data, or in the literature ones for the existence of a cold, fast
rotating NSD, in the core of our Galaxy. We cannot exclude it, either, but what we observe is qualitatively compatible with 
being exclusively due to the CMZ hiding half of the data that are, instead, visible toward other line of sights just outside this region.  \smallskip

\item RV data from the latest release of APOGEE (DR17) do not confirm a kinematic detection of the NSD as clear
as previously reported. The analysis of the updated sample of APOGEE's targets in the inner bulge reveals that the two velocity peaks detected by \citet{schonrich+2015} are much more confused with the rotation of the whole bulge. \smallskip

\end{itemize}

The present work highlights our poor understanding of the Milky Way nuclear region. Strong observational biases affecting this region much more than its surroundings may produce "features" that are not real, but just a consequence of a different selection function.

Deeper photometric data with higher spatial resolution are critical to derive high precision PMs, while a larger sample of stars with a measured RV (and hopefully abundances) are needed to establish how different this region is from its neighbors. While several fiber spectrographs and integral field units will be available soon at the largest observatories, as far as photometry is concerned, only the James Webb Space Telescope has both the spatial resolution and depth to allow us to properly characterize the 3D structure -- and star formation history -- of this unique region of our Galaxy. We emphasize the need for a complete map of the innermost 4 degrees of the Galaxy, as it is crucial to compare what we see inside the inner 2 degrees with what we see just outside them. Moreover, we need to map the spatial variation of the transverse and radial velocities in order to understand whether this is truly a cold, fast rotating disk, or just the inner part of the rotating bulge.

\begin{acknowledgements}
Based on observations taken within the ESO VISTA Public Survey VVV, Program ID 179.B-2002, made public at the ESO
Archive and through the Cambridge Astronomical Survey Unit (CASU). \\

This work is funded by ANID, Millenium Science Initiative, ICN12\_009 awarded to the Millennium Institute of Astrophysics  (M.A.S.), by the ANID BASAL Center for Astrophysics and Associated Technologies (CATA) through grant FB210003, and by  FONDECYT Regular grant No. 1230731. A. R. A. acknowledges support from DICYT through grant 062319RA. E. V. acknowledges the Excellence Cluster ORIGINS Funded by the Deutsche Forschungsgemeinschaft (DFG, German Research Foundation) under Germany’s Excellence Strategy – EXC-2094-390783311. 

\end{acknowledgements}

\bibliographystyle{aa}
\bibliography{biblio} 

\end{document}